\documentclass[letterpaper,twocolumn,10pt]{article}
\usepackage{usenix-2020-09}

\usepackage{amsmath}
\usepackage{amsfonts}
\usepackage{graphicx}
\usepackage{subcaption}
\usepackage{booktabs}
\usepackage{tabularx}
\usepackage{adjustbox}
\usepackage{float}

\usepackage{url}

\usepackage{breakurl}

\usepackage[ruled,vlined,linesnumbered]{algorithm2e}
\renewcommand{\SetProgSty}[1]{\renewcommand{\ProgSty}[1]{\textnormal{\csname#1\endcsname{##1}}\unskip}}
\SetProgSty{textup}
\SetArgSty{textup}

\newcommand{\parb}[1]{\vspace{1pt}\noindent{\bfseries #1.}}
\newcommand{\parbi}[1]{\vspace{1pt}\noindent{\bfseries\itshape #1.}}

\def\Snospace~{\S}

\usepackage{xspace}
\newcommand{\ie}{\emph{i.e.,}\xspace}
\newcommand{\eg}{\emph{e.g.,}\xspace}

\newcommand{\name}{PPipe\xspace}

\newcommand{\comment}[1]{{\em \color{blue} {#1}}}

\newcommand{\jonnycomment}[1]{{\em \color{brown} {(Jonny: #1)}}}
\newcommand{\qiang}[1]{\textcolor{blue}{#1}}
\newcommand{\cut}[1]{{}}

\usepackage[font=small]{caption}

\usepackage{enumitem}
\usepackage{multicol}

\usepackage[available,functional,reproduced]{usenixbadges}

\usepackage{etoolbox}
\newtoggle{entered}

\SetAlgoSkip{algoskip}

\usepackage{fancyhdr}

\begin{document}

\title{\name: Efficient Video Analytics Serving on Heterogeneous GPU Clusters via Pool-Based Pipeline Parallelism}

\author{
{\rm Z. Jonny Kong}$^{*}$\\
Purdue University
\and
{\rm Qiang Xu}$^{*}$\\
Purdue University
\and
{\rm Y. Charlie Hu}\\
Purdue University
}

\maketitle
\fancypagestyle{firstpagefooter}{
  \fancyhf{}
  \fancyfoot[C]{In the Proceedings of the
2025 USENIX Annual Technical Conference, Boston, MA, July 2025}
}
\thispagestyle{firstpagefooter}

\renewcommand{\thefootnote}{\fnsymbol{footnote}}
\footnotetext[1]{Both authors contributed equally to the paper.}

\begin{abstract}

With the rapid innovation of GPUs, heterogeneous GPU clusters in both public
clouds
and on-premise data centers have become increasingly commonplace.
%
%
{In this paper, we demonstrate how pipeline parallelism, a technique
well-studied for throughput-oriented deep learning model training, can be used
effectively for serving latency-bound model inference, \eg in video analytics
systems, on heterogeneous GPU clusters.}
Our work exploits the synergy between diversity in model layers and diversity
in GPU architectures, which results in comparable inference latency for many
layers when running on low-class and high-class GPUs.
We explore how such overlooked capability of low-class GPUs can be exploited
using pipeline  parallelism and present a
novel inference serving system, \name,
that employs {\em pool-based pipeline parallelism}
via an MILP-based control plane
and a data plane that
performs
resource reservation-based adaptive batching.
Evaluation results on diverse workloads (18 CNN models) show that \name
achieves 41.1\%--65.5\% higher utilization of low-class GPUs while maintaining
high utilization of high-class GPUs, leading to 32.2\%--75.1\% higher serving
throughput compared to various baselines.
\end{abstract}

\section{Introduction}
\label{sec:intro}

\begin{figure}[tp]
	\centering
	\includegraphics[width=.8\linewidth,trim=0 10 0 0]{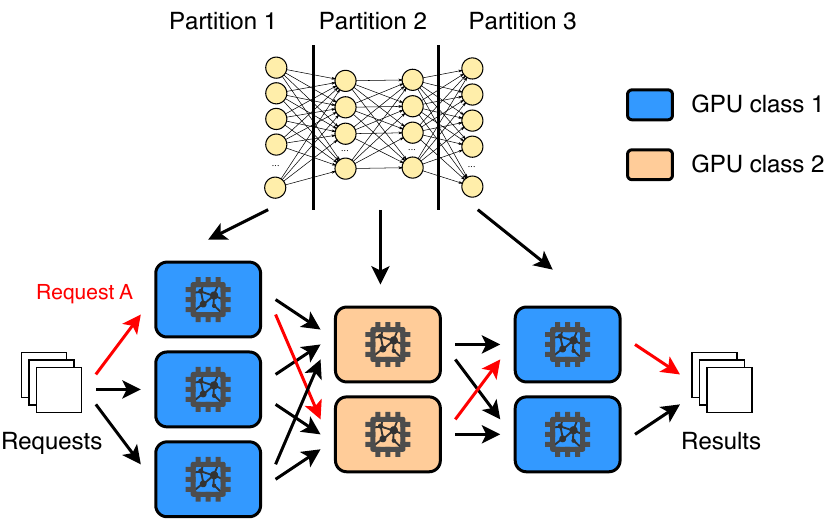}
    \caption{Pool-based pipeline parallelism on a heterogeneous GPU cluster
        with model partitioning. Each request (\eg request A) is processed by
        all partitions sequentially, and may be processed by any of the GPU
        servers allocated to each partition (GPU pool).
    }
	\label{fig:pipeline}
	\vspace{-5mm}
\end{figure}

While applications based on large language models (LLMs) have witnessed
remarkable advancements in recent years, video analytics
systems~\cite{mostcameracitiesinus,beijingcamera,londoncamera,mostcameracitiesintheworld},
which leverage extensive networks of cameras deployed in major cities across
the U.S. and around the world~\cite{beijingcamera, londoncamera,
billionicameras}, remain heavily reliant on traditional machine learning models
such as convolutional neural networks (CNNs).
These systems play a critical role in enabling a wide range of
applications, including real-time surveillance for security purposes,
efficient transportation management through traffic monitoring,
enhanced public safety via crowd analysis, and improvements in
healthcare through patient monitoring.
{The global video analytics market is estimated to grow from \$3.2 billion
in 2023 to \$19.1 billion by 2030, at a CAGR of 29.2\%~\cite{global24}.}

In video analytics systems,
streams of video frames from cameras
deployed at different locations of interest are 
uploaded to the cloud servers that perform analytics, \ie deep neural network
(DNN) model inference.
Serving such inference requests 
is challenging for two primary reasons.
First, video analytics inference requests often have stringent service level
objectives (SLOs), \eg
200~ms~\cite{jiang2018chameleon,zhang2021elf}.
Second, inference requests from real-world applications can be
bursty~\cite{10.1145/3341301.3359658,10.1145/3419111.3421285,285173,288582}.
Meeting the latency SLO for all inference requests
requires provisioning hardware resources for the peak load, which can
be costly as the hardware resource becomes under-utilized during
off-peak periods.

As with model training, model inference relies on the use of accelerators such as GPUs.
With rapid innovation of GPUs~\cite{nvidia_hopper_architecture}, newer
generations of GPUs are introduced to the market in short release cycles.
Yet, their high cost and limited supply have dis-incentivized cloud vendors
and private organizations from retiring older
generations
of GPUs.  As a result,
these organizations are increasingly operating highly heterogeneous GPU
clusters~\cite{weng2022mlaas}.

This paper studies how to serve popular DNN models, \ie with high volumes of
requests~\cite{azurecognitive}, on heterogeneous GPU clusters.
Being able to do so not only allows
utilizing low-class computing resources that are otherwise
unusable~\cite{park2020hetpipe} in clusters dedicated to model serving,
\eg in edge clouds or private clouds running AI-based
apps~\cite{google_distributed_cloud_edge,azure_private_mec},
but also, as we will show in this paper, can
significantly enhance the serving throughput of high-class GPUs.

In particular, we explore the under appreciated benefits of pipeline
parallelism among lower-class and higher-class GPUs in online model serving.
While the benefits of pipeline parallelism
have been well
studied
for throughput-oriented model
training~\cite{park2020hetpipe,kim2016strads,10.1145/3386367.3432728,NEURIPS2022_2b4bfa1c,Ding_Botzer_Weninger_2021},
its potential for model serving under latency (SLO)-constrained settings
on heterogeneous GPU clusters has been largely unexploited.
Intuitively, the benefits of partitioning a model and pipelining the partitioned
inference among mixed high-class and low-class GPUs appear limited.
For example, if the high-class GPU is 10$\times$ faster (\ie lower latency)
than the low-class GPU for a given model,
then simply running 1/10 of the model layers on the low-class GPU already leads
to 1.9$\times$ longer total latency.

We instead make a key observation about the performance characteristics of DNN
model inference on heterogeneous GPUs\cut{ that suggests
lower-class GPUs can be utilized to supplement higher-class GPUs to
achieve higher throughput without compromising their SLOs}:
{\em the two forms of diversity in
model inference on a heterogeneous GPU cluster --- diversity in model layers and
in GPU types --- can interact with each other
synergistically.}
%
First, a DNN model typically has many layers with diverse operations and tensor
dimensions, leading to varying GPU utilization on a given GPU.
%
%
Second, more importantly, for the same DNN model, the relative
per-layer inference latency on different classes of GPUs
can vary significantly across the model layers.
%
%
This observation suggests that partitioning the DNN model in a GPU-aware manner
and executing each partition on the GPU type that runs most effectively 
can improve the effectiveness of all GPU types
and hence the inference throughput of the whole cluster.
%
Effectively, lower-class GPUs
offload part of the model
inference from higher-class GPUs
with minimum elongation of the end-to-end inference latency.

\cut{
How to develop a practical model serving system that exploits
the above observation, however, faces several challenges.
}

While this observation provides guidelines on efficient pipeline parallelism on
heterogeneous GPU servers, partitioning a model and running the
partitions along a {\em chain} of GPUs, as done in pipeline parallel DNN
training~\cite{huang2019gpipe,narayanan2019pipedream}, is too stringent and
leads to suboptimal partitions:
all partition stages must have matching latencies to avoid pipeline stalls,
which restricts the flexibility of model partitioning and leaves less
opportunity for low-class GPUs to run layers they are efficient at.

%
To this end,
we present \name, a model serving system that harnesses mixed GPU
types in heterogeneous GPU clusters
to maximize serving throughput.
\name is built on three key ideas.
First, to improve scheduling flexibility,
it employs {\em pool-based pipeline parallelism} where each
model partition is associated with a {\em pool} of GPU servers, and
each request can be processed by any GPU
in each partition
pool along the pipeline, as shown in \autoref{fig:pipeline}.
{\em Such pooled pipeline parallelism
allows different partitions to have different numbers of GPU
servers, varying inference latencies, and run with different
batch sizes, as long as the inference throughput provided by each pool
of GPU servers matches with each other.
}

Second, to realize the scheduling flexibility exposed by pool-based
pipelined model inference,
\name generates the optimal
configuration of pool-based pipelined model inference,
\eg one that maximizes the 
model serving throughput of a given cluster while meeting inference SLOs, 
using Mixed Integer Linear Programming (MILP), which takes as input
the per-layer inference latency on all candidate GPUs and under all
candidate batch sizes from offline profiling.
\cut{optimally split a DNN model among the inference
pipeline of heterogeneous GPU pools and determine the batch size per
split can be highly complex.
}

Such an MILP-based optimal solution, however, 
effectively assumes ideal request arrivals, \ie synchronous arrival of batches in
locksteps at all GPUs of the first partition pool, and flow down the
pipeline partitions in sync.
In practice, the inference requests
arrive asynchronously and can be
bursty~\cite{10.1145/3341301.3359658,10.1145/3419111.3421285,285173,288582},
which creates transient high load that overwhelms the throughput
prescribed in the MILP solution, introducing several sources of extra delay not
accounted for in the MILP solution and leading to SLO violations.

To bridge the gap between the MILP solution and runtime dynamics due to
asynchronous and bursty request arrivals, 
\name treats the MILP-based formulation as the control plane which prescribes
optimal DNN model partition and GPU allocation,
and employs a novel data plane that performs
{\em resource reservation-based adaptive batching}
to address the unique challenges in batching
pooled-based pipelines: deciding for each batch
{\em which pooled pipeline, which path within the pipeline, and the batch size.}
Our scheduler overcomes the above challenges by
(1) maintaining (current and future) availability of resources (GPUs and
network bandwidth)
in the pooled pipelines; and
(2) probing them
to find the maximal batch of requests that can meet the SLOs of each request
when the batch reaches the end of the pipeline path.

\if
schedules and coordinates dynamically formed
batched inferences across the the pool of GPUs to ensure the
batches injected into the pipelines meet their inference SLOs
in the presence of bursty request arrival.
\fi

We evaluate \name using production workloads on top of 
100-GPU large-scale simulations and 16-GPU testbeds on Google Cloud consisting
of a variety of high- and low-class GPUs such as NVIDIA V100, L4, T4, and P4.
Evaluation across 18 CNN models shows that \name achieves 44.1\%--65.5\% higher
utilization of low-class GPUs while maintaining high utilization of high-class
GPUs compared to various baselines, leading to 32.2\%--75.1\% higher serving
capacity, while successfully processing 99\% of the requests without dropping
or SLO violations.

In summary, we make the following contributions:
\begin{itemize}[nosep,left=0pt]
\item The first exploration of {pipeline-parallel} model serving on
	heterogeneous GPU clusters under latency (SLO)-constrained settings.
  \item 
    The complete design of \name,
    which employs three design ideas to maximize inference throughput of heterogeneous GPU clusters:
    pool-based pipeline parallelism,
    an MILP-based control plane that prescribes optimal pool-based
    pipeline plans,
  and a data plane that performs resource reservation-based adaptive batching to  
  handle runtime dynamics due to asynchronous and bursty request arrivals.
\item
  Extensive evaluation of \name showing \name outperforms baseline designs by
  32.2\%--75.1\% in inference throughput and 41.1\%--65.5\% higher low-class
  GPU utilization.
\end{itemize}

The source code of \name is available at
\url{https://github.com/JonnyKong/PPipe}.

\section{Motivation and Key Idea}
\label{sec:moti}

We motivate how low-class GPUs can be effectively
used to augment high-class GPUs in a heterogeneous cluster
in model serving by exploiting pipeline parallelism.

\begin{figure}
	\centering
	\includegraphics[width=.8\linewidth,trim=0 15 0 0]{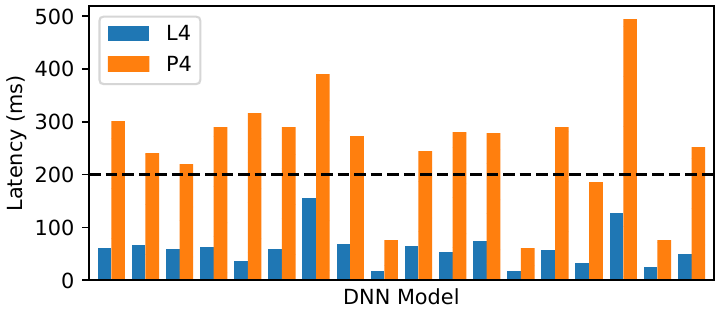}
	\caption{Inference latency of 18 popular DNN models (\autoref{tab:dnn_models})
          under batch size 4 on different GPU classes.}
	\label{fig:inference-latency}
	\vspace{-5mm}
\end{figure}

\parbi{Low-class GPUs fail to meet the inference latency SLO} On a highly
heterogeneous GPU cluster, the inference time on old or low-class GPUs is
usually several times longer than that on newer or high-class GPUs. For the 18
DNN models we use for evaluation (\autoref{subsec:methodology}), the inference
time (on the highly optimized TensorRT~\cite{tensorrt} inference framework) on
the low-class NVIDIA P4 is 3.0x--7.9x longer than that on the high-class NVIDIA
L4, as shown in \autoref{fig:inference-latency}.
Even if
a model runs on the low-class
GPU without violating latency SLO, it can barely perform batched inference,
which could significantly improve GPU utilization and
throughput~\cite{201468,10.1145/3341301.3359658,9355312}.
{As shown in \autoref{fig:inference-latency}, only 22\% of the DNN models
can run on the low-class GPU (NVIDIA P4) at batch size 4} without exceeding
200~ms, a latency SLO target commonly used among video analytics
pipelines~\cite{jiang2018chameleon,zhang2021elf}.

\parbi{Key insight: Diversity in per-layer inference delay across GPUs}
{Given the high latency of low-class GPUs, the benefit of partitioning and
running a model across low- and high-class GPUs, if done in a GPU-oblivious
manner, will be limited.}
For example, if the high-class GPU is 10$\times$ faster (\ie lower latency)
than the low-class GPU for a given model,
then simply running 1/10 of the model layers on the low-class GPU already leads
to 1.9$\times$ longer total latency.
Our key observation is that there exist two forms of diversity in
model inference on a heterogeneous GPU cluster: diversity in model
layers and in GPU types, and they can interact with each other
synergistically.

\begin{figure}[tp]
	\centering
	\includegraphics[width=.55\linewidth,trim=0 15 0 0]{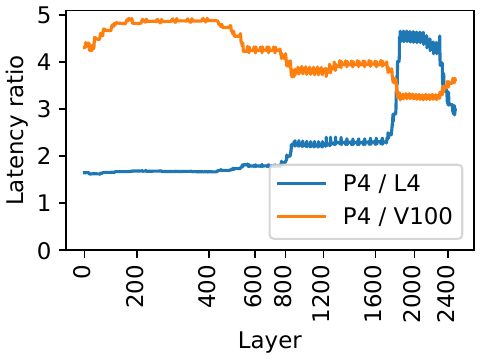}
	\caption{The ratio of inference latency on NVIDIA P4 over L4 and P4 over V100 across EfficientNet-B8~\cite{pmlr-v97-tan19a} layers.}
	\label{fig:latency-ratio}
	\vspace{-5mm}
\end{figure}

In particular, for many popular CNN backbone
architectures, \eg EfficientNet~\cite{pmlr-v97-tan19a} and
ResNet~\cite{He_2016_CVPR}, later layers have more channels compared to earlier
layers, but with lower feature dimensions. Such architectural differences
among layers within a DNN model can lead to different computational
properties on GPU accelerators.
To gain insight into this, we measure the {\em latency ratio},
\ie the ratio of
the inference latency of the same layer on different GPU types for
all the layers within a DNN model. \autoref{fig:latency-ratio} shows that
for EfficientNet-B8 on NVIDIA P4 over L4 and P4 over V100,
respectively, with a moving window of 128 layers. The inference latency ratio on
P4 over L4 is about {1.7}
for early layers, indicating these layers have closer inference
latencies on both P4 and L4. On the other hand, later layers have much
higher latency ratios, and those layers will suffer significant
slowdown running on P4 over L4. If we were to partition the DNN model
and run it on P4 and L4, we should place earlier layers on P4 and
later layers on L4, which provides higher chances to keep the
inference time below the latency SLO and enables batching
opportunities. All 18 DNN models we studied (\autoref{tab:dnn_models}) exhibit
varying latency ratios across layers and we omit the rest due to page
limit.

Interestingly, the latency ratios between P4 and V100 show completely
different trends on EfficientNet, where earlier layers exhibit much higher
latency ratios than later layers. In this case, one would run on P4
later layers instead. Such differences in the trends of latency ratios
happen due to GPU design tradeoffs, architectural improvements,
and their interaction with DNN layers of different
characteristics. For example, GPUs with more SMs or higher ops:bytes
ratio provide more benefits for layers of larger size or higher
arithmetic intensity~\cite{gpuperformance}.

Such varying trends in per-layer latency
ratios on different GPUs suggest that
partitioning a DNN model in a {\em GPU-aware} manner 
is critical in exploiting pipeline parallelism
so that high-class and low-class GPUs can work on model layers
that they are optimized for, which improves their 
efficiency and hence
the inference throughput of the whole cluster.

\parbi{Key idea: pool-based pipeline parallelism}
To apply model partitioning to exploit the diversity of per-layer inference ratio on
different GPUs, a simple approach is
partitioning a model and pipelining the inference of partitions along
a {\em chain} of GPUs similarly as in pipeline parallelism in DNN
training~\cite{huang2019gpipe,narayanan2019pipedream},
and feature maps
generated by one GPU are transferred to the next
downstream GPU. Such {single-chain} inference pipelines have the advantage
of simple scheduling and coordination, but also come with two major
drawbacks: (1) To avoid pipeline stalls, all partitions
need to have
similar inference latencies.
However, such a partitioning strategy is too stringent
and will lead to suboptimal partitions, \eg leaving layers with high
latency ratios (on high/low-class GPUs) to run on the low-class GPU or having a large feature map at
the partition point.  (2) Many GPUs cannot take advantage of
heterogeneous inference when the cluster contains more GPUs of one
class than the other.

To provide more scheduling flexibility, we instead associate each partition
with {\em a pool of GPU servers} of the same type, and each request can be
processed by any GPU allocated to the first partition, and then continue
the inference on any GPU
in the second partition, and so on, as shown in \autoref{fig:pipeline}.
This approach mitigates the drawbacks of {single-chain} pipelines:
different partitions can have different
numbers of GPU servers
(\eg $N_1$ servers in pool 1, $N_2$ servers in pool 2),
have different inference latencies
(\eg $t1$ and $t2$ for the 2 pools),
or even run with different batch sizes 
(\eg $b1$ and $b2$ for the 2 pools),
as long as the
inference throughput provided by each pool of GPU servers
matches well with each other
(\eg $N_1*b_1/t_1 = N_2*b_2/t_2$),
and the total latency is below the latency SLO.
In an optimal partitioning, multiple such {\textit{pooled pipelines}} may be
employed at the same time,
employing different ways of partitioning
the DNN model running on different GPU pools.

\section{Prelude to \name: Basic MILP Formulation}
\label{sec:prelude}


To exploit pool-based pipelined inference, one needs to figure out the optimal
way to partition a DNN model,
the placement of the DNN partitions onto GPU servers, and the batch size for
the GPUs in each partition. It is relatively straightforward to formulate an
MILP problem to figure out the optimal solution. We briefly describe the MILP
formulation below 
and present the full mathematical formulation in Appendix \autoref{sec:milp}.

\parbi{Inputs}
The MILP formulation takes as input the GPU count of each GPU type, the
interconnect bandwidth of the target cluster, and the inference latency SLO. To decide
the optimal model partitions, it also requires
the intermediate feature map sizes and
the inference latencies of
individual layers within a DNN model under different batch sizes and on
different GPU models,
which can be obtained from the profiling output of TensorRT~\cite{tensorrt}.

\parbi{Encoding model partition and placement}
Suppose the GPU cluster consists of 2 GPU types and we restrict a DNN model to
be divided into at most 3 partitions. The placement of DNN model partitions
falls into one of 14 potential pooled pipelines:
if partitioned into 2 partitions, each partition can run on {a pool of}
either GPU types (4 pipeline options);
if partitioned into 3, each of the 3 partitions again has the choice to run on
{a pool of} either GPU types (8 pipelines);
the DNN model can also directly run on a {pool of} either GPU type without
partitioning (2 pipelines).
%
{As discussed in \autoref{sec:moti}, the optimal solution may contain
multiple pooled pipelines.}

For each partition within a pooled pipeline, we need to decide on the exact partition
points, \ie the first and last layers. To this end, we construct a set of
binary decision variables for each partition indicating whether each layer is the
first or last layer of a partition. Apart from that, we also create decision
variables 
to represent the batch size and the number of GPUs used by each
partition.

\parbi{Constraints}
The total latency of each pooled pipeline, including the inference latency of
each partition and the feature map transfer latency between partitions (both
can be derived from the batch size), should be below the latency SLO; the total
GPU count used by all partitions pertaining to a specific GPU type should not
exceed the GPU count for that GPU type.
Finally, the inference throughput of a pooled pipeline is bottlenecked by the
partition of lowest throughput, where each partition's throughput can be
calculated based on its batch size, inference latency, and the number of GPUs
allocated to the partition.
We observe that CNN models commonly used in video analytics pipelines are
typically not memory-constrained on datacenter GPUs; hence, the MILP formulation
does not account for GPU memory.

\parbi{Objective}
By default, we try to maximize the total inference throughput of the GPU
cluster, which is the sum of the throughputs of all pooled pipelines employed by the
MILP solution. The MILP formulation can also be configured for other objectives
like minimum server cost~\cite{10.1145/3472883.3486993} or provisioned
power~\cite{9773234}. In the presence of multiple DNN models, given the ratio
between the DNNs' workloads, the MILP formulation computes the normalized
throughput for each DNN (throughput divided by the DNN's workload percentage),
and maximizes the lowest normalized throughput among the DNNs.

\parbi{Outputs}
The solver of the MILP formulation outputs the
pooled pipelines employed by the optimal plan, \ie those being allocated at least 1 GPU. For
each pipeline, the solver outputs the DNN model partition points, the batch
size used by the GPUs in each partition, and the number of GPUs allocated to each
partition.


{
In essence, MILP holistically determines an optimal set of pooled pipelines by
selecting the model partition points for each pipeline, along with the GPU type,
count, and batch size for each partition in each pipeline, all of which
affect the overall throughput of the GPU cluster.
}


\section{Challenges in Developing a Working System}
\label{sec:challenge}

While the MILP formulation above provides the optimal plan in theory,
turning it into a working DNN serving system 
faces several practical challenges, as discussed below.

\parb{C1: Extensive search space of the MILP formulation} The MILP
formulation needs to decide the first and last layers of a DNN
partition, whose complexity depends on the number of layers in a DNN
model. For the set of representative models in our evaluation
(\autoref{subsec:methodology}), the average layer count is 613.2. The
partition points need to be searched for all partitions across all
pipelines, making the search space combinatorial. The search space is
further inflated by additional dimensions including inference batch
size and GPU count used by each partition. With such a vast search
space, it takes more than 7 hours (running the Gurobi~\cite{gurobi} solver on a
Google Cloud n1-standard-64 instance) to obtain the optimal solution for 80
layers,
making it impractical to adapt to changing workload, \eg diurnal
load~\cite{9773234}.

\parb{C2: Asynchronous and bursty request arrival}
In essence, the MILP formulation outputs a solution
that assumes ideal inference request arrival. Suppose a pipeline solution consists of
two partitions with 40~ms inference latency each, and the inference throughput
is 1000 requests per second. The MILP solution effectively assumes 
that 40 requests arrive at the same time every 40~ms, which are
simultaneously processed by all GPUs allocated to the first partition, and then
forwarded to the second partition, and so on.

In reality, in an online inference system,
inference requests arrive asynchronously and in a bursty manner, which can disrupt
the MILP solution with two forms of extra delays:
(1) Early arriving requests have to wait for later requests to form a batch
to be dispatched to a GPU in the first partition, incurring {\em initial batching delay} (D1);
(2) The staggered batched inference initiated on the GPUs in the first partition
will cascade through the remaining partitions in the pipeline.
In such staggered pipelined inference, it is possible
when a GPU in partition $i$ finishes inference on a batch,
all of the GPUs in partition $i+1$ are still busy running other batches,
causing {\em inter-partition queuing delay} (D2).
Such queuing delay is further complicated when partitions use
different batch sizes, requiring the split and merge of batches
which creates complex dependencies between the GPUs of different partitions.

To incorporate the above extra delays at runtime, we could add a predefined margin to
the latency SLO as input to the MILP
formulation~\cite{10.1145/3341301.3359658,10.1145/3295500.3356164}
which will output
adjusted (still fixed) batch sizes.  But simply adding a
static margin cannot handle bursty request arrival,
which can still result in either too many or too
few transient requests compared to the
{statically}
adjusted target batch size.
Such dynamic conditions require
{\em dynamically adjusting the batch sizes.}
%

\parb{C3: Network contention}
\cut{
On public clouds, network transfer exhibits high tail latency due to network
dynamics like background traffic. For example, our measurement shows that on
GCP, the tail transfer latency is about 5x the theoretical transfer latency,
which leads to a third form of extra delay (D3) not captured by the MILP
solution.

On the other hand, in practice,
}%
{
%
We observe that heterogeneous clusters such as Google Cloud Platform (GCP) and
Amazon Web Services (AWS) come with high-bandwidth networks that theoretically
can finish the transfer of a feature map with a small percentage of the total
inference latency.
For example, GCP's P4 instances have a bandwidth of 32~Gbps, which can
theoretically transfer feature maps of CenterNet (with batch size 1) which range
from 3~MB to 50~MB in 0.8--13.2~ms.
%
%
However, it is common for multiple GPUs to
collocate on the same server, sharing the server's network bandwidth.
This can lead to network contention between GPUs (D3) that
disrupts pipeline schedules and causes SLO violations.
}
%
%
The contention becomes more severe when we divide each GPU into multiple
virtual GPUs (\autoref{subsec:bs-unify}), 
which increases the number of ``GPUs'' on the same server that likely transfer
feature maps at the same time. Note that this issue cannot be addressed by
conservatively allocating each virtual GPU an equal share of the available
bandwidth, as it results in low bandwidth for all virtual GPUs, significantly
limiting the benefit of pipeline parallelism.

\section{\name Design}

\begin{figure}[tp]
	\includegraphics[width=\columnwidth,trim=0 10 0 0]{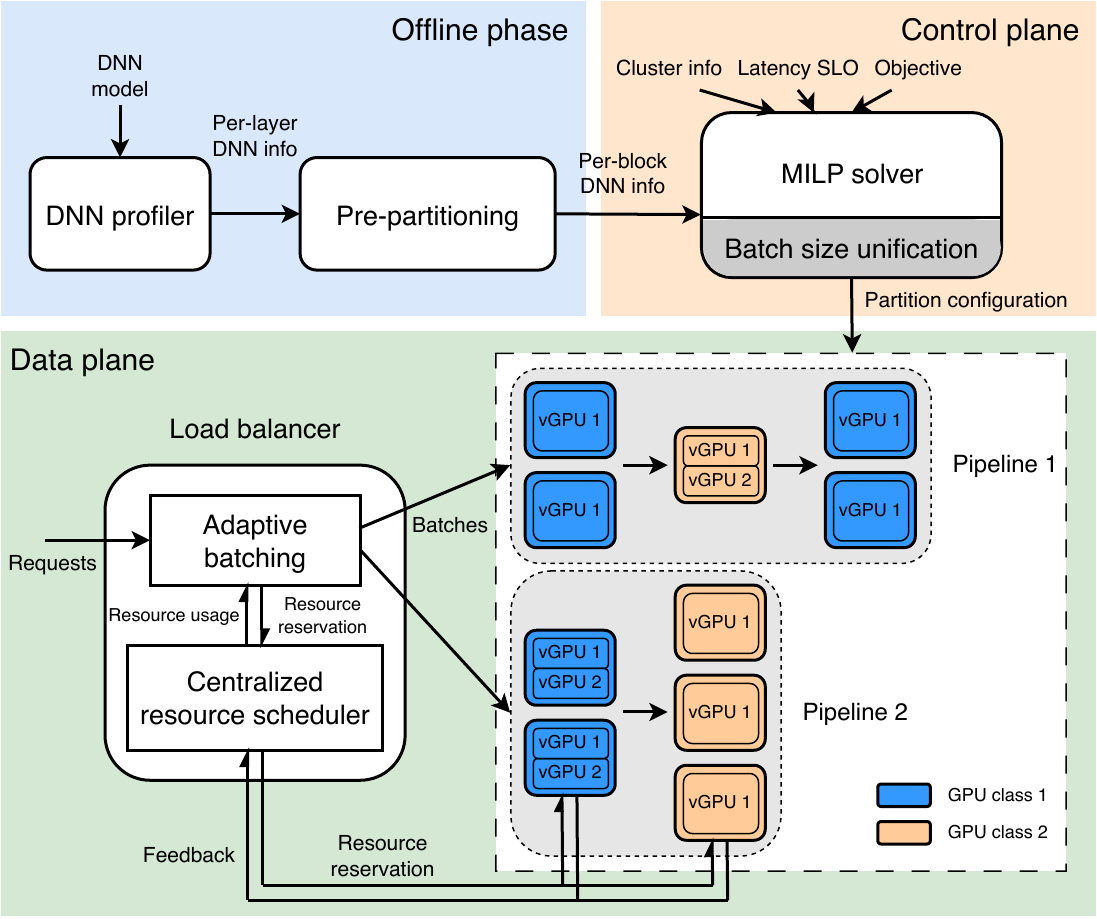}
	\caption{\name architecture.}
	\label{fig:arch}
	\vspace{-5mm}
\end{figure}

As discussed above, designing a practical pool-based pipeline parallel inference
serving system faces a key challenge: the MILP formulation does not
capture or handle runtime dynamics due to asynchronous and bursty request
arrival or delayed feature map transfer from network contention.
\cut{
  (1) the MILP formulation does not capture the three sources of extra delays
in the inference pipeline at runtime;
(2) it does not prescribe an effective {\em data plane}
that handles bursty request arrivals at runtime.
}%
We tackle these challenges by splitting \name, our {\em pool-based} pipeline parallel
inference serving system for heterogeneous GPU clusters, into a control plane
and a data plane.
First, \name treats the MILP-based formulation as the control plane that
prescribes optimal DNN model partitions and GPU allocation
{for each pooled pipeline}.
Second, to handle
delays caused by asynchronous and bursty request arrivals (D1 \& D2) and network contention
(D3), \name employs a novel data plane that
performs {\em resource reservation-based adaptive batching} to ensure the request
batches injected into the pooled pipeline meet their latency SLOs.

\subsection{Architecture Overview}
\label{subsec:arch}

\autoref{fig:arch} shows the architecture of \name, which 
consists of an offline phase, a control plane, and a data plane.

\parbi{Offline phase}
In the offline phase, the DNN model profiler profiles a model's per-layer
inference latency on different GPU types.
To reduce the search space of the MILP solver (C1), we design a
pre-partitioning method that groups the layers of each DNN model into blocks
which are fed into the MILP solver (\autoref{subsec:pre-part}). With this
method, the MILP solver only needs to find partition points among a few blocks
instead of hundreds of layers, significantly reducing the search space.
We profile each model and each block independently, and each block is profiled
on every GPU type and batch size.
The profiling is fast, requiring only a few hours to cover all 18 DNN models.
Since the addition of models happens infrequently, and each model will be
served over a long time, the one-time offline profiling cost is amortized and
manageable.
When solving MILP, the latency of each partition is computed as the sum of the
latencies of its constituent blocks.
 
\parbi{Control plane}
The control plane, which runs the MILP solver, takes as input the profiling
information of DNN blocks and the inference latency SLO for each DNN, the
cluster information (GPU count for each GPU model), along with high-level
objective, \eg maximum throughput, and outputs the partitioning of each DNN model
and allocation of GPU resources across partitions in each pooled pipeline (detailed in
\autoref{sec:prelude}).
We observe that the
synchronization among partitions ({\bf C2}) could be much simpler if partitions
within the same pooled pipeline all use the same batch size. As such, we
enhance the MILP formulation with {\em batch size unification}.
%
\cut{
Furthermore, \name deducts a predefined margin from the request latency SLO to
provision batch building time under bursty request arrivals,
and discount the
inter-GPU bandwidth by a scaling factor to accommodate the tail latency in
network transfers, which we observe is about 5x the theoretical transfer
latency on public clouds.
}

The MILP solver runs periodically, triggered dynamically in response to
workload changes, such as shifts in the load ratio when serving multiple DNNs,
which typically occur once every one or a few
hours~\cite{9773234,zhong2024distserve}.
%
Note that the MILP solver runs on the CPU, and is asynchronous to the
inferences running on the GPUs.
Migrating to a new MILP plan involves reassigning GPUs to different partitions
or pooled pipelines, and having the data plane dispatch the requests according
to the updated plan.
To minimize migration latency, each GPU asynchronously preloads the new model
weights into memory ahead of the switch, without interrupting the ongoing
inference of the existing model; this is feasible because GPU memory is
generally not a bottleneck for vision models.
Once all GPUs complete loading, \name pauses ingesting new requests to perform
a pipeline flush, which takes 
about 1x the SLO of the currently serving DNNs (in the order of 100s of
milliseconds).
After the flush, all GPU switch to the new weights simultaneously and the data
plane resumes request dispatching.
In essence, each migration incurs a downtime of several hundred milliseconds,
which is negligible compared to the interval between migrations.

\parbi{Data plane}
The data plane groups inference requests into batches and executes
them through the
pools of GPUs
in each of the pipelines prescribed by MILP.
%
{To address the extra delays D1--D3, we
design a novel adaptive batching scheme that
selects the pooled pipeline to execute the next batch, one GPU from each pool to
run the corresponding model partition (the \textit{pipeline path}), and the actual
batch size that meets the request SLOs.}

\subsection{DNN Pre-Partitioning}
\label{subsec:pre-part}

\cut{
\begin{figure}
	\subcaptionbox{EfficientNet-B8~\cite{pmlr-v97-tan19a}.\label{fig:feature-size-boundary-efficientnet}}
	{\includegraphics[width=.49\linewidth,trim=0 15 0 0]{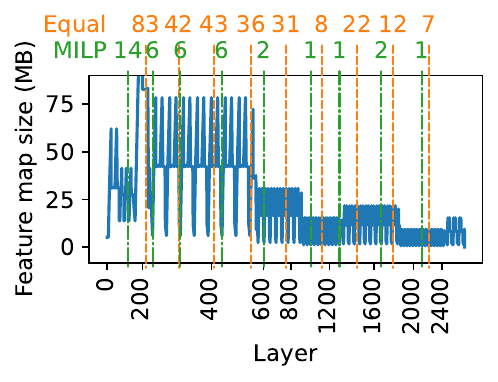}}
	\hfill
	\subcaptionbox{EfficientDet-D1~\cite{tan2020efficientdet}.}
	{\includegraphics[width=.49\linewidth,trim=0 15 0 0]{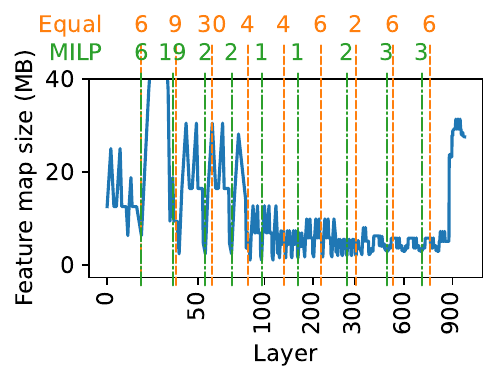}}
	\vspace{-2mm}
	\caption{Feature map sizes of two DNN models. The $x$-axis is scaled
		according to the per-layer inference latency. Vertical lines show the
		boundaries and boundary sizes (on top) of the DNN blocks generated by
		the naive
		approach (Equal) and our pre-partitioning approach (MILP). Our approach
		provides smaller 
          feature map sizes at block boundaries.}
	\label{fig:feature-size-boundary}
	\vspace{-5mm}
\end{figure}
}

As discussed in \textbf{C1}, the sheer number of layers within a DNN model
results in a huge search space for the MILP formulation and prevents the solver
from finding good plans in a reasonable amount of time.
To this end, we devised a simple DNN pre-partitioning approach that groups the
layers in a DNN model into a few ($N$) blocks of approximately equal runtime
{on a selected GPU type};
empirical results show that the choice of GPU type has minimal impact on the partitioning.
%
%
%
Specifically, we start from the first layer and sequentially group
consecutively layers together until their combined runtime is as close as
possible to $1/N$ of the runtime of the entire DNN; this process is repeated
until we reach the last layer.
After grouping layers into blocks, we profile the blocks on different GPU types
and with different batch sizes, as needed by the MILP solver. It takes only less
than 10 minutes to profile a single DNN model, as each block can be profiled
independently; when solving MILP, we obtain the latency of each partition by
adding up the latencies of the blocks in it.

With pre-partitioning, the MILP solver only needs to find partition points
among the $N$ blocks (we tried $N=5$ to $N=20$ and found $N=10$ provides a good
balance between plan optimality and MILP running time), instead of 613.2 layers
on average across the set of models
(\autoref{sec:eval}).
As a result, the MILP runtime is significantly reduced to 3.5 seconds on average
over different DNN model and GPU cluster setups.

\cut{
\parbi{Complexity of \name's MILP}
\comment{
Is it a function of N, number of GPus, ????
In practive, the MILP solver runs in ???, as shown in Section????.
In particular, it scales well with the number of GPUs in the cluster, as ????.
}
}

\subsection{Batch Size Unification}
\label{subsec:bs-unify}

We tackle the key challenge of the data plane --- bridging the gap between MILP
solution and runtime dynamics (\textbf{C2} and \textbf{C3}) --- in two steps. In the first step,
we simplify the challenges faced by data plane scheduling
with {\em batch size unification}.

As discussed in \textbf{C2}, mismatch of batch sizes between partitions
within a pooled pipeline requires batches to be merged and split
which complicates scheduling of batched inference across partitions.
%
Things can be substantially simplified if all partitions
within the same pooled pipeline use the same batch size. However, as the GPUs have
different computational capacities, in the plans generated by the MILP planner,
high-class GPUs tend to use larger batch sizes compared to low-class GPUs to
improve GPU utilization and inference throughput.
Naively forcing all partitions within the same pooled pipeline to use a
uniform batch size can lead to under-utilization of high-class GPUs
and degrade overall system performance.

Instead of
forcing all partitions of a pooled pipeline to use the same batch size,
we vary the GPU size to equalize the batches per GPU.
%
Specifically, we
incorporate virtual GPUs
into the MILP
formulation, so that the MILP solver can take into account the throughput and
inference latency differences between batches on different partitions and make
holistic decisions in choosing a unified batch size.
To this end, instead of feeding a GPU as a whole to MILP, we
feed four possible virtual GPU types: 1/1, 1/2, 1/3 and 1/4 of a physical GPU
(this is achieved with Multi-Process Service (MPS)~\cite{mps} during runtime).
The use of virtual GPUs only mildly expands the search space of the MILP solver
as there are only 4 virtual GPU types. Further, we profile the per-block
inference latencies under not only different batch sizes and GPU types, but also
different virtual GPU types.\footnote{We capture the interference between
virtual GPUs during profiling by running the same DNN on all virtual GPUs of the
same physical GPU in parallel.} Finally, we
add additional
constraints to the MILP formulation requiring all partitions within the same
pooled pipeline to use the same batch size.
The mathematical representation of the enhanced MILP formulation is provided in
Appendix \autoref{sec:milp-mps}.

\subsection{Resource Reservation-Based Adaptive Batching}
\label{subsec:sched}

With batch size unification, \name stills needs to perform adaptive
batching at runtime, \ie dynamically forming and scheduling batches,
in serving asynchronous and bursty inference requests.

Compared to pipelined inference over a single chain of GPUs,
dynamic batching in pool-based pipelines
faces unique new challenges.
In a chain of GPUs, the scheduler just needs to
find the largest batch size that satisfy the end-to-end SLO.
In a cluster of pooled pipelines (output by the MILP solution),
the batching scheduler has to make three decisions
for each batch: {\em which pooled pipeline, which path within the pipeline,
and the batch size.}
The decisions are further complicated by:
(1) the optimal batch size depends on which pipeline the batch is sent to;
(2) the optimal pipeline path in turn depends on the batch size
and resource availability in the pooled pipeline.

%

\parbi{Resource reservation-based adaptive batching}
Our resource-reservation-based adaptive batching
scheduler overcomes the above challenges by
(1) maintaining (current and future) availability of resources (GPUs and network)
in the pooled pipelines; and
(2) probing them
to find the maximal batch of requests that can meet the SLOs of each request
when the batch reaches the end of the pipeline path.

\begin{figure}[tp]
    \centering
    \begin{subfigure}[t]{0.47\columnwidth}
        \includegraphics[width=\textwidth,trim=0 0 0 0]{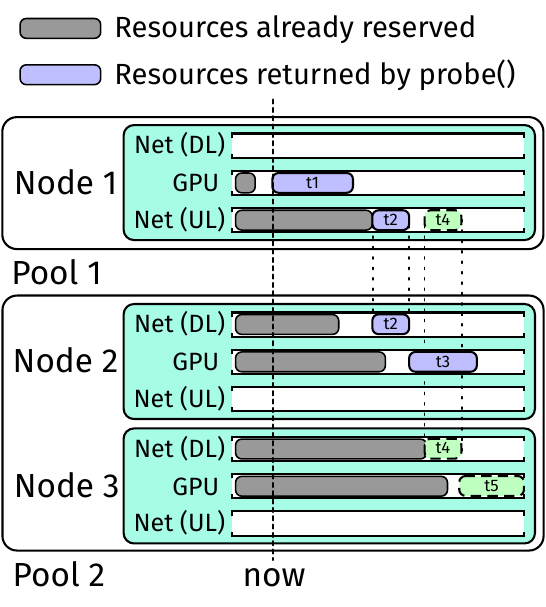}
        \vspace{-6mm}
        \caption{
            \texttt{probe()} takes as input a candidate pooled pipeline and
            batch size, and outputs         
            the optimal pipeline path and the intervals
            of resources to be used (in blue).
        }
        \label{fig:probe_before}
    \end{subfigure}
    \hfill
    \begin{subfigure}[t]{0.47\columnwidth}
        \includegraphics[width=\textwidth,trim=0 0 0 0]{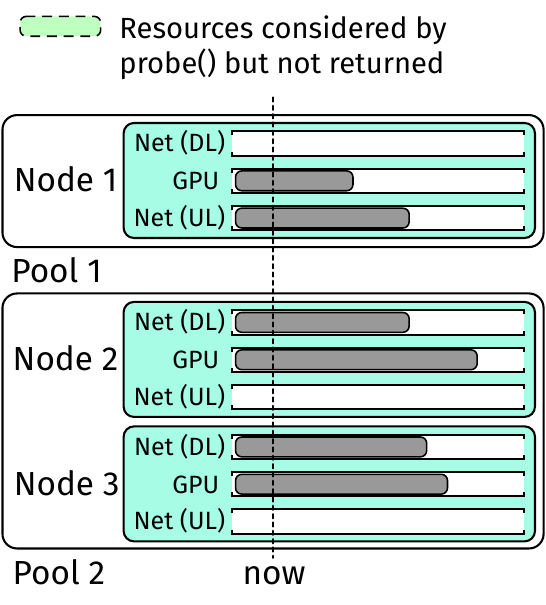}
        \vspace{-6mm}
        \caption{\texttt{reserve()} marks the resource intervals as reserved
        until the end of the intervals returned by \texttt{probe()}.}
        \label{fig:probe_after}
    \end{subfigure}
    \vspace{-2mm}
    \caption{
        The resource reservation mechanism on an example pooled pipeline
        comprising two partitions, with 1 and 2 GPUs, respectively.
    }
    \label{fig:probe}
    \vspace{-4mm}
\end{figure}

The scheduler works in two steps, both using a stateless probing procedure
\texttt{probe()}.
The procedure takes a specific pooled pipeline
and a hypothetical batch size as input, and outputs the pipeline path that
minimizes the end-to-end (E2E) inference time under current resource
availability, {with an example shown in \autoref{fig:probe_before}.}
The E2E inference time includes the per-stage inference time, network transfer
time, and the waiting time for required resources (GPU and NIC) along the
pipeline path.
The detailed algorithm of \texttt{probe()} is explained later.

In Step 1, we identify the pooled pipeline $i$ that can complete a batched
inference at the pipeline's unified batch size $bs_i$ (as determined by the
MILP solution) with the shortest waiting time under the current resource
availability.
This is achieved by invoking \texttt{probe()} for each pooled pipeline with its
corresponding unified batch size, and selecting the pipeline that has the
lowest waiting time,
where the waiting time is computed as the sum of delays waiting for each
required resource along the pipeline path.
{
Using waiting time as the metric effectively balances the load between the pipelines because
it is a good indication of the load of the pipeline, \ie the lower the
waiting time, the lower the load on the pipeline.
%
}

{
The E2E inference time of sending a batch of size $bs_i$ down the selected
pipeline $i$ in Step 1, however, may not meet the SLO (as the batch size
$bs_i$ generated by MILP assumed synchronized request arrival).
}
%
In Step 2, using the chosen pooled-pipeline $i$, we search for {the
actual largest batch size that can meet the SLOs and the corresponding pipeline
path.}
%
Specifically, we {iteratively} invoke \texttt{probe()} with progressively
smaller batch sizes, starting from $bs_i$ (the batch size from the MILP
solution), until the completion time of the pipeline path returned by
\texttt{probe()} falls within the deadline of the first (oldest) pending
request.
Finally, one of three actions is taken depending on the chosen batch size and
the number of pending requests:
(1) If the deadline cannot be met even with batch
size 1, the oldest request will be dropped and the adaptive batching process
starts over from choosing a pooled pipeline (Step 1);
(2) if the number of pending requests is smaller than the chosen batch
size, the batching engine
waits for more requests (till the last moment when the
requests in queue can still be processed without SLO violation);
%
(3) otherwise, the resources
of the selected pipeline path
returned by \texttt{probe()} are
{\em reserved} (by calling \texttt{reserve()}), \ie their availability are
updated {as shown in \autoref{fig:probe_after}}, and the requests at the
head of queue are grouped into a batch of the chosen batch size and dispatched
to the selected pipeline path according to the resource reservation.
Since the batch size is based upon the actual remaining time of the requests in
the queue, the extra delay D1 (initial batching delay) is taken into account,
and the requests in the batch are guaranteed to meet their SLOs.

The pseudocode for our adaptive batching algorithm is provided in Appendix
\autoref{subsec:batch}.

\parbi{The probing procedure}
%
Since adaptive batching works at the batch level, the probing procedure
needs to be fast to keep up with the request arrival rate.
To this end, we implement \texttt{probe()} based on a greedy algorithm.
%
In a nutshell, 
\texttt{probe()} works by sequentially selecting one GPU from each GPU pool
along the pooled pipeline.
For each GPU pool,
it goes through all GPUs allocated to the pool
and selects
the GPU that minimize the completion time for that partition stage.
During this process, \texttt{probe()} determines which resources—GPUs, uplink
bandwidth, and downlink bandwidth—need to be reserved for specific future time
intervals.
Note that since feature map transfer require simultaneous availability of
network resources on both the sending and receiving sides, \texttt{probe()}
ensures the allocation of network resources that satisfy both the uplink of the
preceding GPU and the downlink of the current GPU.

We illustrate the workflow of \texttt{probe()} with an example in
\autoref{fig:probe_before}.
Consider a pooled pipeline consisting of two pools, containing one and two GPUs
respectively.
For simplicity, we assume each GPU resides on a separate node, although in
practice, multiple GPUs may reside on the same node and share the same network
resources.
In the first pool, \texttt{probe()} selects node 1 as it is the only option,
which requires reserving its GPU for time duration \texttt{t1}.
In the second pool, \texttt{probe()} needs to decide between nodes 2 and 3.
Selecting node 2 requires the simultaneous reservation (as shown by the
vertical dotted line) of node 1's uplink and node 2's downlink network during
\texttt{t2}, as well as node 2's GPU during \texttt{t3}, after which the
request completes; alternatively, selecting node 3 results in a completion time
by the end of \texttt{t5}.
Since node 2 results in an earlier completion time than node 3,
\texttt{probe()} selects node 2. In the end, it returns the
selected pipeline path with associated resource 
allocated: node 1's GPU during \texttt{t1}, node 1's uplink and node 2's
downlink network during \texttt{t2}, and node 2's GPU during \texttt{t3}.
%
%

Since \texttt{probe()} is based on real-time resource availability, it directly
takes into account extra delays D2 (inter-partition queuing delay) and prevents
D3 (network contention delay).
Furthermore, \texttt{probe()} runs in real time, and scales linearly with the
number of virtual GPUs allocated to the pooled pipeline.
The pseudocode for \texttt{probe()} and \texttt{reserve()} are provided in
Appendix \autoref{subsec:batch}.

\parbi{Feedback correction}
The scheduler maintains resource reservation tables that keep track of when
each resource will be used. However, this scheduler's view of resource usage
might deviate from reality due to variations in inference time and network
bandwidth.
To this end, we let all nodes report back to the scheduler when the
reserved resources were actually used immediately after every resource usage.
The scheduler updates the resource usage table accordingly. The feedback
correction mechanism ensures that the scheduler's view of resource usage is
synchronized with reality at all times.

\parbi{Extra SLO margin in the control plane}
While our resource reservation-based adaptive batching algorithm ensures
requests meet their SLOs, we notice that the resulting batch size may be much
smaller than that in the MILP output due to extra delays D1--D3, causing 
{large deviations from the MILP plan}. 
To bridge the gap between control plane planning and data plane execution, we
deduct an empirically determined margin from the SLO when running the control
plane MILP solver, so that the adaptive scheduler picks the same batch sizes as
in the MILP output most of the time.

\parbi{Dispatching Complexity}
In the worst case, dispatching a batch requires a number of \texttt{probe()}
function calls equal to the product of the number of pipelines in the cluster
and the number of candidate batch sizes.
Each \texttt{probe()} function call has a time complexity linear to the number
of GPUs within the corresponding pipeline.
Consequently, the adaptive batching algorithm incurs  low runtime overhead, as we
will demonstrate in \autoref{subsec:end_to_end_results}.

\cut{
\subsection{Discussions}

\qiang{Both questions are out of context?}

\parbi{How does \name's resource reservation scheme make the time
  spent between partitions more predictable?}
Our resource reservation scheme effectively maintains a global view
of when each GPU and NIC will be idle or occupied in the near
future. With such information, the scheduler can accurately predict
how long a new batch of requests needs to wait before resources for
the next partition become available, thus making the time spent
between partitions as well as the total latency more predictable.

\parbi{Extending \name to accommodate tensor parallelism}
Tensor parallelism can be easily incorporated in IPIPE as it is
actually the opposite of supporting virtual GPUs. While virtual GPUs
divide a single GPU to run multiple workers, tensor parallelism
combines multiple GPUs for a single worker. Thus, to support tensor
parallelism in IPIPE, we simply need to go through a similar process
as supporting virtual GPUs, i.e., profile the DNN models on different
tensor parallelism configurations, and extend the MILP by adding a
dimension of tensor parallelism size.
}

\section{Implementation}

\parbi{Offline phase and control plane}
We implement the offline phase and control plane in Python in 2.7~kLOC. We
use Gurobi~\cite{gurobi} as the MILP solver. We work with DNN models in their
ONNX format and TensorRT format interchangeably, since the ONNX format
provides flexibility, while the TensorRT format provides high inference
performance.

\parbi{Data plane}
We implement both a discrete-event simulator for modeling large-scale
GPU clusters and a prototype implementation for \name's data plane, in
about 9.0~kLOC.
The simulator is written in Java and maintains a global event queue sorted by
timestamp, executing events in chronological order.
Supported event types include request arrivals, batch dispatches, per-partition
executions, and feature map transfers, etc.
At each simulation step, the simulator dequeues the next event, invokes the
corresponding event handler,
which updates the system states and may produce additional events to be added
to the event queue (\eg a request arrival event adds the request to the waiting
queue and triggers a scheduler event).

\cut{
The real-world implementation is implemented in a combination of Julia and C++,
with the inter-node communication implemented with TCP for control messages
and NVIDIA NCCL for feature maps.
}
The prototype implementation is written in a combination of Julia and C++,
using TCP for control message exchanges and NVIDIA NCCL for transferring
feature maps between nodes.
To minimize the feature map transfer latency, we quantize float32 feature maps
to float16 (only at partition boundaries), effectively reducing the transfer
size by half. We find such quantization has negligible impact on task accuracy.
The accuracy dropped by 0.00\%, 0.01\%, and 0.01\% for object recognition,
object detection, and instance segmentation tasks, respectively.

\section{Evaluation}
\label{sec:eval}

In this section, we evaluate \name's serving performance under a variety of DNN
models from different tasks, considering various combinations of low-class and
high-class GPUs.
We show that \name can serve 32.2\%--75.1\%
more requests compared to various
baselines while meeting 99\% SLO attainment, on the discrete-event
simulator with 100 GPUs.
Additionally, on 16-GPU clusters deployed on Google Cloud, \name achieves
16.7\%--52.8\% higher serving throughput.
We also conduct sensitivity analysis to show the impact of GPU
composition and SLO on \name's performance.

\subsection{Methodology}
\label{subsec:methodology}

\begin{table}[]
\centering
\caption{Heterogeneous Cluster (HC) setups.}
\label{tab:heterogeneous_cluster_setups}
\vspace{-3mm}
\small
\resizebox{\columnwidth}{!}{
\begin{tabular}{ll|cll|c}
\toprule
\multicolumn{1}{c}{Setup} & \multicolumn{1}{c|}{GPUs} & Setup & \multicolumn{1}{c}{Instances} & \multicolumn{1}{c|}{GPUs} & BW (Gbps) \\ \midrule
HC1-L & \begin{tabular}[c]{@{}l@{}}25$\times$ L4,\\ 75$\times$ P4\end{tabular} & HC1-S & \begin{tabular}[c]{@{}l@{}}4$\times$ \texttt{g2-standard-16},\\ 2$\times$ \texttt{n1-highcpu-16}\end{tabular} & \begin{tabular}[c]{@{}l@{}}4$\times$ L4,\\ 12$\times$ P4\end{tabular} & 50 \\ \hline
HC2-L & \begin{tabular}[c]{@{}l@{}}25$\times$ L4,\\ 75$\times$ T4\end{tabular} & HC2-S & \begin{tabular}[c]{@{}l@{}}1$\times$ \texttt{g2-standard-48},\\ 6$\times$ \texttt{n1-highcpu-32}\end{tabular} & \begin{tabular}[c]{@{}l@{}}4$\times$ L4,\\ 12$\times$ T4\end{tabular} & 32 \\ \hline
HC3-L & \begin{tabular}[c]{@{}l@{}}25$\times$ V100,\\ 75$\times$ P4\end{tabular} & HC3-S & \begin{tabular}[c]{@{}l@{}}2$\times$ \texttt{n1-highcpu-16},\\ 12$\times$ \texttt{n1-highcpu-16}\end{tabular} & \begin{tabular}[c]{@{}l@{}}4$\times$ V100,\\ 12$\times$ P4\end{tabular} & 50 \\ \hline
HC4-L & \begin{tabular}[c]{@{}l@{}}25$\times$ V100,\\ 75$\times$ T4\end{tabular} & HC4-S & \begin{tabular}[c]{@{}l@{}}1$\times$ \texttt{n1-standard-64},\\ 6$\times$ \texttt{n1-highcpu-32}\end{tabular} & \begin{tabular}[c]{@{}l@{}}4$\times$ V100,\\ 12$\times$ T4\end{tabular} & 32 \\ \bottomrule
\end{tabular}
}
\vspace{-2mm}
\end{table}

\begin{table}[]
\centering
\caption{DNN models used in the evaluation.}
\vspace{-3mm}
\label{tab:dnn_models}
\resizebox{\columnwidth}{!}{
\begin{tabular}{l|l|l|l}
\toprule
\multicolumn{1}{c|}{Recognition} & \multicolumn{1}{c|}{Detection} & \multicolumn{1}{c|}{Segmentation} & \multicolumn{1}{c}{Others} \\ \midrule
ConvNext~\cite{liu2022convnet} & ATSS~\cite{zhang2020bridging} & APCNet~\cite{he2019adaptive} & Color-v2~\cite{zhang2016colorful} \\
EfficientNet~\cite{pmlr-v97-tan19a} & CenterNet~\cite{duan2019centernet} & DNL-Net~\cite{yin2020disentangled} &  \\
GoogleNet~\cite{szegedy2015going} & FSAF~\cite{zhu2019feature} & EncNet~\cite{zhang2018context} &  \\
RepVGG~\cite{ding2021repvgg} & GFL~\cite{li2020generalized} & FCN~\cite{long2015fully} &  \\
WideResNet~\cite{zagoruyko2016wide} & RTMDet~\cite{lyu2022rtmdet} & GCNet~\cite{cao2019gcnet} &  \\
 & EfficientDet~\cite{tan2020efficientdet} & NonLocalNet~\cite{wang2018non} &  \\ \bottomrule
\end{tabular}
}
\vspace{-4mm}
\end{table}

\begin{figure*}[tp]
	\centering
	\includegraphics[width=1.0\textwidth,trim=0 10 0 0]{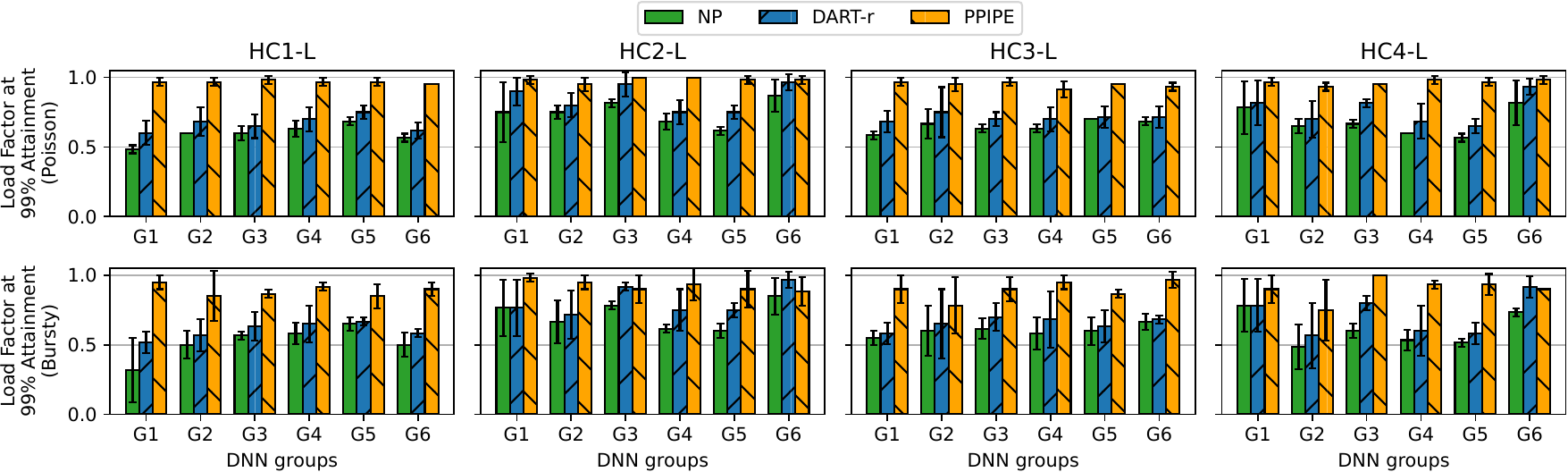}
    \caption{
        Maximum load factor of each framework under 99\% SLO attainment on
        100-GPU clusters, under ``Poisson'' and ``Bursty'' workloads.
    }
    \label{fig:main_results_gain_barplot}
    \vspace{-2mm}
\end{figure*}

\parbi{Cluster configuration}
We consider 4 heterogeneous cluster setups, labelled as HC1$-$HC4 in
\autoref{tab:heterogeneous_cluster_setups}. Each setup consists of a large (L)
100-GPU variant used for the discrete-event simulator, and a small (S) 16-GPU
variant deployed on Google Cloud.
%
Note that a Google Cloud VM instance can host multiple GPUs, resulting in each
HC having a varying number of VMs while maintaining a consistent number of
GPUs. 
Accounting for the scarcity of high-class GPUs~\cite{weng2022mlaas}, our
default configuration includes 25 high- and 75 low-class GPUs for each HC's
large variant, and 4 and 12 for the small variant. We further evaluate \name's
performance under different ratios of high- and low-class GPUs in
\autoref{subsec:sensitivity}.
Note that for GPU-equipped VMs, Google Cloud provisions network bandwidth based
on the number and type of its GPUs, leading to different interconnect
bandwidths across HCs.
{%
Furthermore, the effective bandwidth for both large and small clusters is
only 1/5 the claimed values in \autoref{tab:heterogeneous_cluster_setups} to
accommodate the observed 5$\times$ network tail latency on Google Cloud.
}

\parbi{Workloads}
Following prior works \cite{285173,288582}, we use Microsoft's Azure Function
(MAF) traces from 2019~\cite{shahrad2020serverless} and
2021~\cite{zhang2021faster}, which were originally derived from Azure
serverless function invocations, as representative inference serving workloads.
When serving multiple DNNs in parallel (\autoref{subsec:end_to_end_results}),
functions are assigned to DNNs in a round-robin manner to determine the
workload ratio among DNNs.
The MAF 2019 trace only includes per-minute aggregated request counts for each
serverless function, and thus we issue requests using Poisson arrival at the
given target load.
Conversely, the MAF 2021 trace includes per-request arrival timestamps, and
thus we upscale the trace to the target load and issue requests accordingly.
The Poisson-emulated 2019 trace is less bursty than the 2021 trace, and thus we
refer to the two traces as ``Poisson'' and ``Bursty'', respectively.

\parbi{DNN models}
We select 18 popular DNN models from public DNN registries such as
TorchVision~\cite{torchvision}, OpenMMLab~\cite{openmmlab}, and OpenVINO model
zoo~\cite{openvino}. The selected DNNs serve a variety of popular computer
vision tasks, as shown in \autoref{tab:dnn_models}.

\parbi{Metrics}
We employ two key metrics to evaluate the inference serving capability. First,
\textit{SLO attainment} represents the percentage of requests that are
successfully processed without being dropped or violating the SLO, under a
specific offered load. Second, we measure the maximum load that the system can
handle at 99\% SLO attainment.

\parbi{Baselines}
%
{
%
The large amount of recent works on model serving on heterogeneous clusters do not incorporate
pipeline parallelism, and thus we abstract them into a state-of-the-art
baseline, denoted as NP below.
We also compare \name with the only prior work 
that exploits pipeline
parallelism across heterogeneous GPUs,
DART~\cite{xiang2019pipelined}, which uses a single-chain-based  pipeline of
GPUs. To isolate the benefit of pool-based pipeline parallelism, we enhance all
baselines to use \name's data plane, \ie the resource reservation-based
adaptive batching (\autoref{subsec:sched}) in presenting the overall results
(\autoref{subsec:end_to_end_results} \& \autoref{subsec:testbed}).
We then evaluate the benefit of \name's second novel design,
resource-reservation-based adaptive batching, in \autoref{subsec:ablation}.
}
    

{
    \begin{itemize}[nosep,left=0pt]
    \item \textbf{No-Partitioning (NP)}. NP executes the entire DNN on either
      high-class or low-class GPUs without partitioning.
      This way of serving
        DNNs on a heterogeneous cluster is representative of various prior
        works~\cite{234998,273804,10.1145/3458817.3476168,9773234,li2023kairos,10.1145/3472883.3486993,10.1145/3419111.3421285,kannan2019grandslam}.
      In particular, the allocation is done by solving \name's MILP formulation
        without model partitioning.
      When integrated with
      \name's reservation-based adaptive batching, NP effectively
      dispatches the largest possible batch to the next available
        GPU while meeting the SLO for each request in that batch.
\item \textbf{DART-r}. DART~\cite{xiang2019pipelined} is an inference
        framework that partitions a DNN onto heterogeneous CPU and GPU
        cores. However, vanilla DART
        constructs a pipeline by chaining all available GPUs,
        each serving one model partition.
        %
        This restricts its use to small clusters, as long pipelines incur
        significant overhead due to frequent feature map transfers.
        To address this limitation, we introduce DART-r, a modified version
        that replicates DART configurations for
        pairs of low-
        and high-class GPUs (more efficient than longer pipelines from fewer
        feature map transfers).
        If one GPU class has more GPUs than the other, the leftover GPUs are
        allocated to individually run entire DNNs without partitioning.
        \cut{DART-r uses an MILP 
          solver in place of the dynamic programming algorithm in DART. }
\end{itemize}
  }

\begin{figure*}[tp]
	\centering
	\includegraphics[width=1.0\textwidth,trim=0 10 0 0]{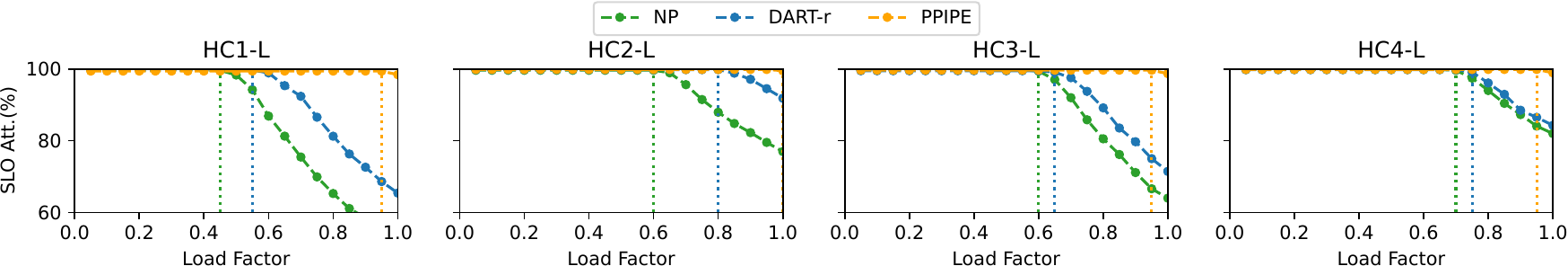}
    \caption{
        SLO attainment of DNN group G1 under the ``Poisson'' trace with varying
        load factors, averaged over the 3 DNNs in the group. The vertical
        dotted line denotes the load factor at which each system reaches 99\%
        SLO attainment.
    }
\label{fig:main_results_attainment_curve}
    \vspace{-2mm}
\end{figure*}

\parbi{Setup}
Following prior work~\cite{288582}, for each DNN, we set the default SLO to be
5$\times$ its inference latency on the fastest GPU (NVIDIA L4) at batch size 1,
resulting in SLOs ranging from 23.4 ms to 165.6 ms.
This reflects the assumption that end users can tolerate higher latency for
more complex models but are indifferent to the underlying GPU used.
We further evaluate \name under other SLOs in \autoref{subsec:sensitivity},
ranging from 2$\times$ to 10$\times$.
As mentioned in \autoref{subsec:sched}, to bridge the gap between control plane
planning and data plane execution caused by the extra delays, a 40\% margin is
deducted from DART-r and \name's MILP formulation, and in NP
in picking the maximum batch sizes that satisfy the SLOs.
When comparing \name with the baselines, we use {\em load factor} 1.0 to denote the
throughput in the output of \name's MILP.
We generate requests using ``Poisson'' and ``Bursty'' traces
respectively, with an average request rate ($\lambda$) ranging from 0.05 to
1.0 times the load factor, at an interval of 0.05. For each $\lambda$, the
experiment lasts 30 seconds.

\subsection{End-to-end Results}
\label{subsec:end_to_end_results}

\parbi{Overall results}
In this section, we evaluate \name's capability of serving DNNs over 100-GPU
clusters (HC1-L to HC4-L) on the discrete-event simulator.
We randomly divide the 18 DNNs into 6 groups of 3 DNNs each (G1--G6), and
serve DNNs within each group in parallel.
%
%
During runtime, we record the maximum load ratio that each DNN can achieve under
99\%
attainment.

\autoref{fig:main_results_gain_barplot} shows under cluster configurations
HC1-L to HC4-L, the maximum load factor achieved under 99\% SLO
attainment, averaged over the DNNs in each group.
First, we find that \name consistently outperforms NP.
\name achieves on average 48.0\% higher load factors than NP on the ``Poisson''
trace (64.9\%, 34.8\%, 46.7\%, and 45.5\% higher for HC1-L to HC4-L, respectively),
and 75.1\% on the ``Bursty'' trace (161.6\%, 34.1\%, 50.4\%, and 54.1\% higher),
showing the advantage of \name's pipeline parallel inference scheme.
Second, compared to DART-r, \name achieves 32.2\% higher load factors on the
``Poisson'' trace (47.2\%, 17.3\%, 34.8\%, and 29.3\% higher on HC1-L to HC4-L,
respectively), and 35.8\%  on the ``Bursty'' trace (50.4\%, 18.1\%, 40.2\%, and
34.5\% higher), showing the advantage of \name's pool-based pipelined inference over
DART-r's chain-based pipelined inference.
%
%
%
Finally, while all frameworks suffer reduced load factors under the ``Bursty''
trace (61.1\%, 69.4\%, and 90.3\% for NP, DART-r, and \name respectively),
compared to ``Poisson'' (66.8\%, 74.9\%, and 96.5\%), the improvement of \name
over the two baselines remain high, showing \name's robustness to varying
request arrival patterns.

\begin{figure}[tp]
	\centering
	\includegraphics[width=0.8\columnwidth,trim=0 12 0 0]{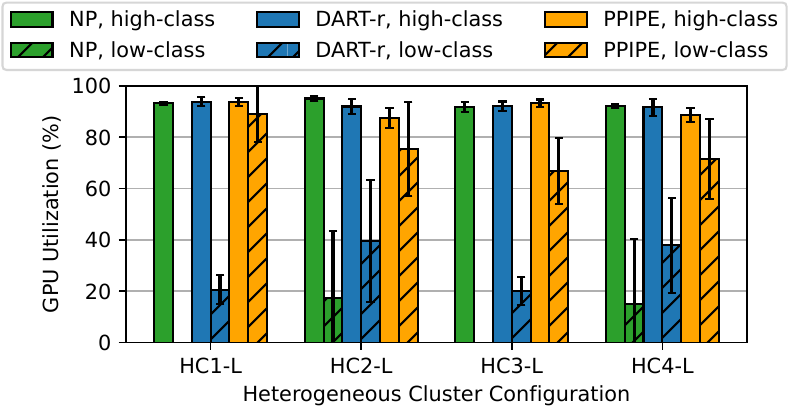}
    \caption{
        GPU temporal utilization under 99\% SLO attainment, averaged over DNNs
        for each cluster configuration.
    }
    \label{fig:main_results_gpu_temporal_util_barplot}
    \vspace{-5mm}
\end{figure}

\autoref{fig:main_results_gpu_temporal_util_barplot} shows each framework's
temporal utilization of high- and low-class GPUs.
For brevity, we only show the utilization on the ``Poisson'' trace.
While all frameworks achieve high utilization of high-class GPUs, NP and DART-r
show zero or low utilization of low-class GPUs.
On average, NP, DART-r and \name achieve low-class GPU utilizations of 8.1\%,
29.5\%, and 73.6\% respectively.
NP's low utilization is caused by the inference time of whole DNNs on low-class
GPUs often higher than the SLO,
prohibiting low-class GPUs from being used.
While DART-r employs DNN partitioning, it chains only one low-end GPU with each
high-end GPU, resulting in under-utilization of excess low-end GPUs when
their number exceeds that of high-end GPUs.

\parbi{SLO attainment under varying load factors}
\autoref{fig:main_results_attainment_curve} shows the SLO attainment under
varying load factors. For brevity, we show only the SLO attainment for DNN
group G1, which includes EfficientNet-B8, EncNet, and RtmDet, and only under
the ``Poisson'' trace.
The attainment is averaged over the 3 DNNs, \eg
on HC1-L at load factor 1.0, \name achieve SLO attainments of 99.3\%, 97.6\%,
and 98.4\% on the 3 DNNs respectively, resulting in an averaged attainment of
98.4\%, which is shown in the figure at load factor of 1.0.
\cut{
For brevity, we showcase only the results
for HC1-L and only one DNN from each task, specifically, the DNN on which \name
achieves the highest gain over NP.
}

We observe that \name outperforms both NP
and DART-r, achieving the highest SLO attainment under the same load factors.
Consequently, it achieves a higher load factor while ensuring 99\% SLO
attainment (\autoref{fig:main_results_gain_barplot}).
For example, on HC1-L, \name achieves 99\% SLO attainment under load factors up
to 0.95, meaning it can handle at least 95\% of the load calculated by the MILP
solver.
In contrast, NP and DART-r's SLO attainment dips below 99\% as the load factor
exceeds 0.45 and 0.55, respectively. This is due to the fact that a load factor
of 1.0 represents the serving capacity of \name, which is higher than the
capacity of NP or DART-r, leading to the dropping of requests beyond their
respective serving capacities.
%
Note that although \name achieves higher load factors than the baselines, it
may not reach the full load factor of 1.0, as seen with HC1-L, HC3-L, and
HC4-L.
This is due to unpredictable request arrival patterns (\eg Poisson), which
cannot be fully accounted for by the MILP solver, as discussed in
\autoref{sec:challenge}.

\parbi{MILP runtime}
\name's MILP solver takes 3.5 seconds on average, showing the effectiveness of
\name's pre-partitioning (\autoref{subsec:pre-part}) in reducing the MILP
complexity.
This latency is negligible compared to the intervals between
re-running the MILP --- triggered by fluctuations in incoming load which occur
relatively infrequently, \eg once every hour~\cite{9773234,zhong2024distserve}.

\parbi{Adaptive batching overhead}
\name's resource-reservation-based adaptive batching mechanism is
lightweight: on a 100-GPU cluster, dispatching a batch requires an
average of only 3.58 \texttt{probe()} calls, incurring a total
overhead of less than 9 microseconds.
This overhead is negligible compared to the batch execution latency, even under
bursty request patterns.

\subsection{Testbed Results}
\label{subsec:testbed}

\begin{figure}[tp]
	\begin{minipage}[b]{.6\linewidth}
        \includegraphics[width=\textwidth,trim=0 12 0 0]{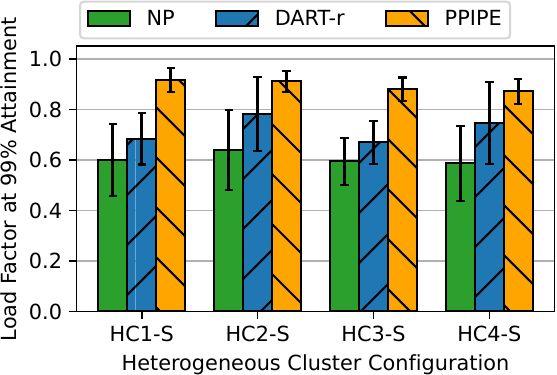}
            \caption{Maximum load factor each cluster configuration can
            achieve under 99\% SLO attainment on the testbed, averaged over the DNNs.}
        \label{fig:testbed_results_gain_barplot}
	\end{minipage}
	\hfill
	\begin{minipage}[b]{.35\linewidth}
        \includegraphics[width=0.78\textwidth,trim=-10 10 10 -1]{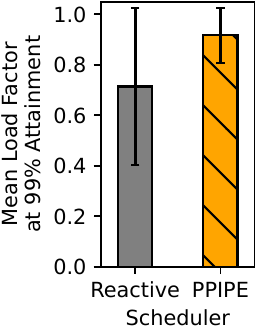}
        \caption{
            SLO attainment with the reactive scheduler and \name's data plane
            scheduler.
        }
        \label{fig:ablation}
	\end{minipage}
    \vspace{-2mm}
\end{figure}

We verify \name's DNN serving capability on
real-world 16-GPU heterogeneous cluster testbeds (HC1-S to HC4-S) deployed on
Google Cloud.
Due to the testbed's smaller GPU counts, instead of serving DNNs in groups of
three as in \autoref{subsec:end_to_end_results}, we serve one DNN at a time
with Poisson-arrival requests.
\autoref{fig:testbed_results_gain_barplot} shows the maximum load factor under
99\% SLO attainment averaged over the 18 DNNs.

First, we observe that \name consistently outperforms NP, achieving
42.6\%--52.8\% higher load factors at 99\% SLO attainment across the cluster
configurations.
This validates the advantage of \name's model-parallelism inference
in a real-world setting.
Secondly, compared to DART-r, \name achieves 16.7\%--34.1\% higher load
factors, which shows the advantage of \name's pool-based pipelined inference.
In summary, \name's significant performance gains over NP and DART-r,
previously demonstrated in the discrete-event simulator, remain
consistent when deployed on a real-world testbed.

\subsection{Ablation Study: Benefit of Resource Reservation}
\label{subsec:ablation}

As discussed in \autoref{subsec:sched}, \name's resource reservation-based data
plane dynamically schedules requests onto GPUs to overcome the delays caused by
bursty request arrival and network contention.
%
%
To demonstrate its effectiveness, we compare it to a \textit{reactive},
distributed adaptive batching scheduler.
%
%
The reactive scheduler performs adaptive batching independently for each GPU
pool in a pooled pipeline.
For each pool, it selects the largest possible batch size that meets the pool's
SLO (as determined by the MILP solver).
%
A similar idea was used in previous model-granularity pipeline scheduling in
\cite{10.1145/3341301.3359658}.

\cut{
we compare it to the distributed data plane design in
Nexus~\cite{10.1145/3341301.3359658}.
}

\cut{
We implement both data plane designs in a discrete-event simulator, which
allows us to scale the cluster setup to 100 GPUs (25 L4 and 75 T4), which
provides more scheduling opportunities, \eg network contentions can be more
flexibly resolved with more GPUs in a partition's pool, to showcase the benefit
of the data plane scheduler.
}

\autoref{fig:ablation} shows the maximum load factor achieved by the two data
plane designs under 99\% SLO attainment on HC2-L, averaged over the DNNs, under
the ``Poisson'' arrival.
\name achieves an average load factor of 0.92, while reactive only achieves
0.71.
The primary factor contributing to this performance degradation is
that the reactive, distributed scheduler lacks resource usage tracking which leads to
batches being scheduled onto servers with saturated network links, causing
bloated transfer delays.
For example, for EfficientNet-B8, where \name achieves 99\% SLO attainment at
1.0 load factor, the reactive scheduler only manages a 0.35 load factor.
With 10~Gbps {effective} bandwidth, feature map transfer between the first and
second partition should take 5.1~ms, but the reactive scheduler leads to
transfer times of 18.9~ms on average and 35.7~ms at the 99\% percentile,
resulting in excessive request drops at the second partition in order for the
remaining requests to meet their SLOs.

\cut{
Secondly, the reactive scheduler may suffer from load imbalance within
pipelines due to its lack of global resource awareness. Batches are distributed
to the pipeline without the knowledge of downstream partition saturation,
leading to request drops at overloaded partitions.
For instance, in one of the MILP-generated pipelines for WideResNet, the first
partition is able to process 4\% higher throughput than the second partition.
Without global knowledge of the resource usage, batches are distributed to the
pipeline despite significant overload and request drops at the second partition.
In summary, \name's global resource scheduler is essential to coordinating
the partitions and ensuring that requests meet their latency SLOs.
}


\subsection{Microscopic Analysis}

\begin{figure}[tp]
	\centering
	\includegraphics[width=0.75\columnwidth,trim=0 15 0 0]{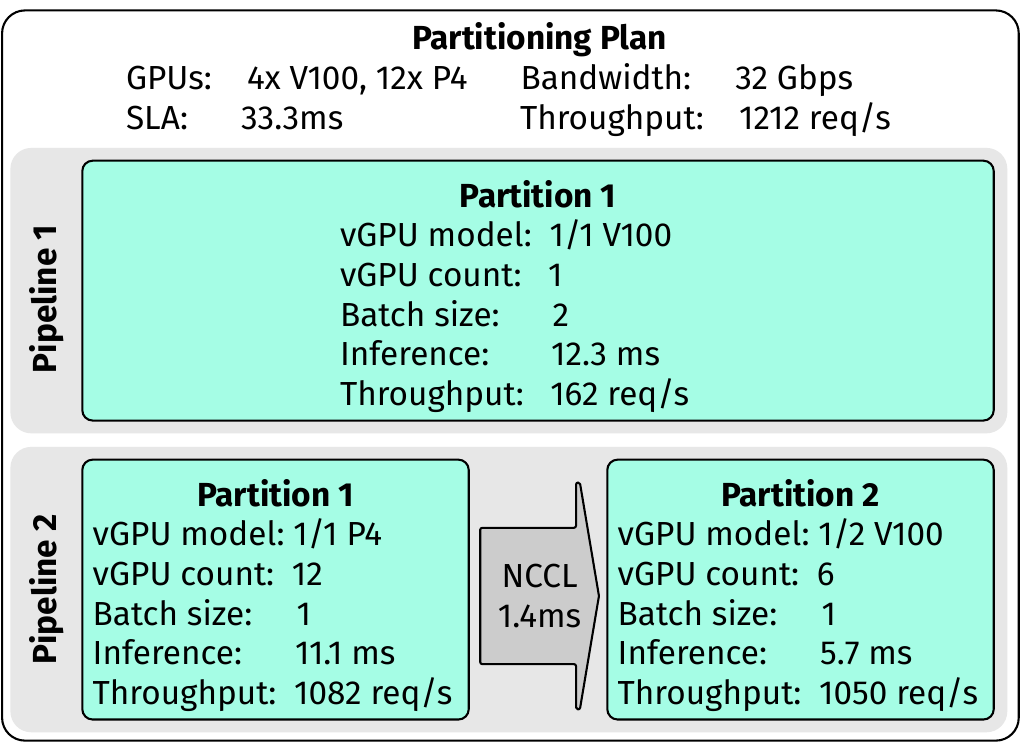}
    \caption{Partitioning plan for the FCN model on HC3-S.}
    \label{fig:microscopic_analysis_partitioning_plan}
    \vspace{-2mm}
\end{figure}

\begin{figure}[tp]
	\centering
	\includegraphics[width=0.78\columnwidth,trim=0 10 0 0]{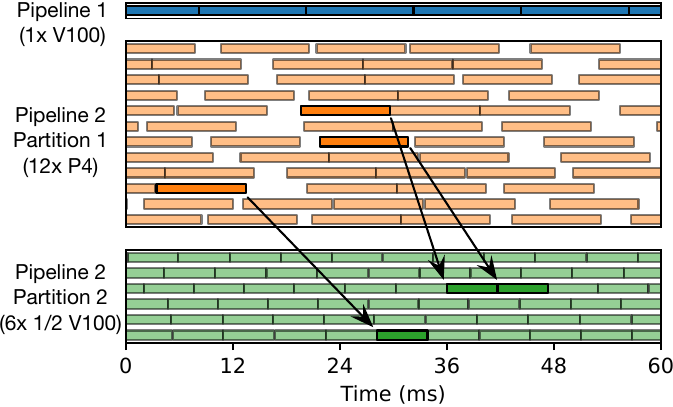}
    \caption{
        An example timeline serving the FCN model on HC3. Each row represents a
        vGPU, and each box corresponds to a batched inference. The highlighted
        pairs of boxes denote the same batches across the partitions within a pooled
        pipeline.
    }
    \label{fig:microscopic_analysis_timeline}
    \vspace{-2mm}
\end{figure}

In this section, we provide a microscopic analysis of \name, using the FCN model
served on the HC3-S cluster testbed on Google Cloud as an example, where \name achieves
a load factor of 0.95 under 99\% SLO attainment.

\parbi{Plan structure}
\autoref{fig:microscopic_analysis_partitioning_plan} shows the partitioning
plan generated by \name's MILP solver for cluster HC3-S, which consists of
4$\times$ V100 and 12$\times$ P4 GPUs. The inference latency of the FCN model
on the fastest GPU (NVIDIA L4) is 6.66 ms, establishing an SLO of 33.3 ms under
an SLO scale of 5. The resulting plan comprises two pipelines, one with a
single partition and the other with two partitions.

The first pipeline consists of a single V100 GPU, performing inference with a
batch size of 2, where each batched inference takes 12.3~ms. The
theoretical throughput of this pipeline is $(2 \times 1 / 0.0123)=162$ requests
per second.
The second pipeline consists of two partitions, with 12 P4 and 3 V100 GPUs
respectively, with 1.4~ms of feature map transfer time in between. Employing
batch size unification (\autoref{subsec:bs-unify}), both partitions perform
batch size 1 inference.
To achieve such a unified batch size, \name divides the three V100s in the
second partition into six virtual GPUs using NVIDIA MPS.
Furthermore, we observe that the two partitions yield similar throughputs of
1082 and 1050 requests per second respectively, resulting in a total throughput
of 1050 (the minimum of the two). This shows \name's ability to balance
resource allocation between partitions to achieve matched throughput.

\parbi{Runtime behavior}
\autoref{fig:microscopic_analysis_timeline} shows an example timeline of the
DNN inference on each virtual GPU. We observe that \name performs inference
back-to-back in pipeline 1, as well as pipeline 2 partition 2, fully using
their GPUs. Note that pipeline 2 partition 1 experiences underutilization,
due to the fact that it was provisioned with slightly higher serving throughput
than partition 2
(\autoref{fig:microscopic_analysis_partitioning_plan}). Furthermore, the figure
showcases that \name's pool-based pipeline allows a batch to be processed by
any GPU within each partition, allowing different partitions
to use different numbers of GPUs
to accommodate different per-partition inference latencies.

\subsection{Sensitivity Analysis}
\label{subsec:sensitivity}


\begin{figure}[tp]
    \centering
    \begin{subfigure}[b]{.44\columnwidth}
        \includegraphics[width=\columnwidth, trim=0 5 -5 0]{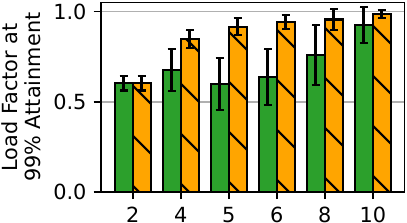}
        \caption{SLO scale.}
        \label{fig:ablation_varying_slo_lineplot}
    \end{subfigure}
    \begin{subfigure}[b]{.27\columnwidth}
        \includegraphics[width=\columnwidth, trim=1 5 1 0]{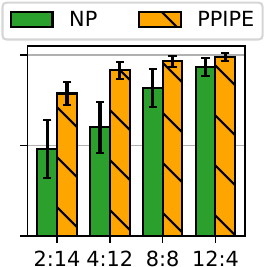}
        \caption{GPU ratio.}
        \label{fig:ablation_varying_gpu_ratio_lineplot}
    \end{subfigure}
    \begin{subfigure}[b]{.26\columnwidth}
        \includegraphics[width=\columnwidth, trim=-10 5 -20 0]{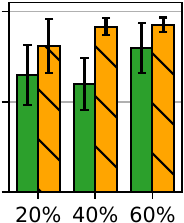}
        \caption{MILP margin.}
        \label{fig:ablation_varying_sla_discount_barplot}
    \end{subfigure}
    \vspace{-2mm}
    \caption{Sensitivity of \name to various factors. Results are averaged over
    18 DNNs on HC1-S.}
    \vspace{-4mm}
\end{figure}

In this section, we study \name's sensitivity to various factors, on the cluster
HC1-S testbed deployed on Google Cloud.

\parbi{Varying SLO scales}
Our main evaluation, which follows AlpaServe~\cite{288582}, uses 5x the
inference latency as the SLO and shows significant performance improvement.
We further evaluate \name considering various SLO scales ranging from 2x to
10x, as shown in \autoref{fig:ablation_varying_slo_lineplot}.
Although 2x and 10x SLOs are less used in practice, we include them to
illustrate and validate \name's expected behavior relative to the baselines.
With an SLO scale of 2, \name shows no improvement over NP, as
such stringent SLOs render the utilization of low-class GPUs impractical for
either NP or \name. Consequently, \name resorts to running entire DNNs on
high-class GPUs, essentially falling back to NP. Conversely, as the SLO scale
increases to 10, \name's improvement over NP becomes marginal, due to the fact
that more DNNs can now meet the SLO running entirely on low-class GPUs, and
hence the low-class GPUs can be utilized in NP, thereby narrowing the gap
between NP and \name.

\parbi{Varying GPU ratios}
\cut{Up to this point, our evaluation has covered cluster configurations HC1-HC4,
each comprising of 4 high-class and 12 low-class GPUs. }In
\autoref{fig:ablation_varying_gpu_ratio_lineplot}, we evaluate \name under
varying ratios of high-class (NVIDIA L4) to low-class (NVIDIA P4) GPUs, which
shows that \name attains more improvements over NP on clusters with fewer
high-class GPUs. For instance, with a high-low ratio of 2:14, \name achieves a
64.03\% higher load factor; as the percentage of high-class GPUs increases,
\name's improvement diminishes, reaching 5.64\% at a high-low ratio of 12:4.

\parbi{Varying SLO margin size}
As discussed in \autoref{subsec:methodology}, a 40\% margin was subtracted from
the SLO in the MILP formulation of both \name and
NP. The impact of the margin size is two-fold --- a larger margin size
reduces the ideal-case serving capacity (\ie what load factor 1.0 signifies),
but increases the load factor achievable under 99\% SLO attainment in practice.
\autoref{fig:ablation_varying_sla_discount_barplot}
shows that under
varying margin sizes,
\name's load factor
increases with larger
margin sizes, but plateaus as the margin size increases beyond 40\%.
Furthermore, \name achieves the highest gain of 52.8\%
over NP under the margin size of 40\%, but also maintains a relatively high
improvement over NP at 20\% and 60\% margin sizes, of 24.9\% and 16.4\% respectively.

\cut{
\subsection{Serving Multiple DNNs Simultaneously}

\begin{figure}[tp]
	\centering
	\includegraphics[width=0.9\columnwidth,trim=0 5 0 0]{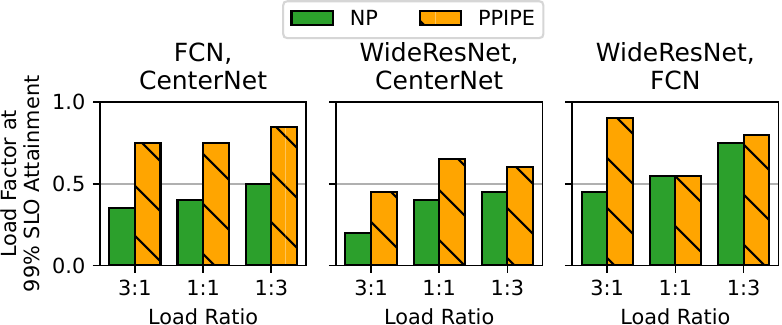}
    \caption{The maximum load factor each system can achieve under 99\% SLO
    attainment, when serving two DNNs simultaneously, under different load
    ratios between the two DNNs.}
    \label{fig:multidnn_barplot}
    \vspace{-5mm}
\end{figure}

\name inherently supports concurrent serving of multiple DNNs on the same
cluster. Given multiple DNNs and the anticipated average request rate ratios
among them, \name's control plane utilizes the MILP solver to generate a
partitioning plan that maximizes the minimum throughputs of each DNN normalized
by the DNN's request rate, ensuring an optimized and efficient distribution of
GPU resources. In the data plane, each DNN runs independently from each other
and operates its exclusive resource scheduler.

We evaluate \name's concurrent serving capability on cluster configuration HC1.
For bervity, we focus on only three pairs of DNNs. For each DNN pair, we generate
concurrent Poisson-arrival requests for two DNNs, at 3:1, 1:1, and 1:3 load
ratios respectively.
%
\autoref{fig:multidnn_barplot} show that \name consistently achieves
high improvments over NP over all DNN combinations, at 90.6\%, 73.6\%, and
35.6\% respectively. Additionally, in a majority of the cases, \name consistently
achieves high improvement across varying load ratios between DNNs. For
instance, when concurrently serving FCN and CenterNet, \name achieves 114.3\%,
87.5\%, and 70.0\% higher load factors than NP under the three load ratios,
respectively. 
}

\begin{figure}[tp]
    \begin{subfigure}[b]{.4\columnwidth}
        \includegraphics[width=\columnwidth, trim=-10 5 -5 -10]{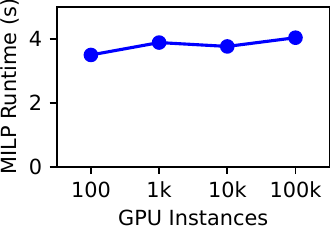}
        \caption{Varying GPU instances.}
        \label{fig:ilp_runtime_increasing_gpu_count}
    \end{subfigure}
    \hspace{10pt}
    \begin{subfigure}[b]{.4\columnwidth}
        \includegraphics[width=\columnwidth, trim=-10 5 -5 -10]{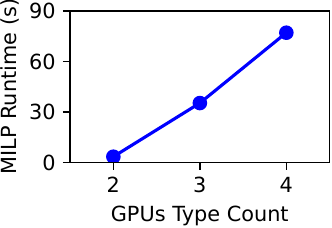}
        \caption{Varying GPU types.}
        \label{fig:ilp_runtime_increasing_gpu_type}
    \end{subfigure}
    \vspace{-2mm}
    \caption{Scalability of \name's MILP-based control plane.}
    \label{fig:ilp_runtime_increasing_gpu}
    \vspace{-4mm}
\end{figure}

\parbi{Varying GPU types and counts}
We analyze how the runtime of \name's MILP-based control plane scales with 
the number of GPU instances and types in a cluster.
First, we scale up the 
HC1-L cluster (\autoref{tab:heterogeneous_cluster_setups})
from 100 to 100k GPUs.
\autoref{fig:ilp_runtime_increasing_gpu_count} shows
that the MILP runtime remains nearly
constant.
This is because adding more GPU instances does not introduce additional MILP
variables, and the problem's complexity stays unchanged.
Second, as the GPU types increase to 3 and 4,
\autoref{fig:ilp_runtime_increasing_gpu_type} shows the MILP runtime increases
to 35.3 and 77.0 seconds,
indicating the number of GPU types has a higher impact on MILP runtime.
However, the increased runtime remains insignificant compared to the interval
between MILP re-executions, which occurs on an hourly
scale~\cite{9773234,zhong2024distserve}.

\section{Related Work}

\parbi{Pipelined partitioned DNN serving}
Several works on DNN serving exploit pipeline parallelism for different
scenarios and objectives.
%
AlpaServe~\cite{288582} improves serving throughput by employing
pipeline parallelism to facilitate multiplexing of GPUs across multiple
DNNs, but it does not take advantage of heterogeneous GPUs.
Another line of works specialize in serving RNN or transformer-based
models~\cite{fang2021turbotransformers,yu2022orca} in a pipelined fashion.
These works do not consider heterogeneous GPU clusters either.
{
Finally, DART~\cite{xiang2019pipelined} partitions DNNs onto a chain of CPUs
and GPUs. However, it does not scale to large GPU clusters, as it will generate
as many partitions as the number of GPUs in the cluster, resulting in frequent
transfer of feature maps which is costly.
}

\parbi{Whole DNN serving on heterogeneous clusters}
Such works typically focus on serving multiple DNNs concurrently, and
study the placement of whole DNNs on servers with the goal of either minimizing the
cost of cloud VMs~\cite{234998,273804,10.1145/3458817.3476168}, or
maximizing the throughput of on-premise clusters~\cite{9773234,li2023kairos}.
Furthermore, several works study serving video analytics applications
that utilize
multiple DNNs forming a Directed Acyclic Graph
(DAG)~\cite{10.1145/3472883.3486993,10.1145/3419111.3421285,kannan2019grandslam}.
None of above works exploits 
the diversity across the layers within a DNN.

\parbi{Whole DNN serving, homogeneous clusters}
Several works focus on model serving on homogeneous
clusters~\cite{288582,10.1145/3341301.3359658,201468,285173,gujarati2020serving}.
%
%
However, these works lack resource allocation mechanisms that leverage DNN
partitioning or GPU heterogeneity.
%
%

\parbi{Model parallel DNN training}
Various works exploit tensor or pipeline parallelism (or both) in model
training~\cite{10.1145/3386367.3432728,NEURIPS2022_2b4bfa1c,Ding_Botzer_Weninger_2021,kim2016strads,huang2019gpipe,narayanan2019pipedream,park2020hetpipe,zheng2022alpa,um2024metis}.
Compared to inference, training has a different set of scenarios and
requirements, \eg
no need to meet SLOs or tackle non-deterministic
request arrivals.

\section{Conclusion}
\vspace{-0.5mm}

In this paper, we presented \name, a system for making effective use
of mixed GPUs on heterogeneous clusters
in serving video analytics applications.
The key innovation of \name is three-fold: pool-based pipelined model
inference, an MILP-based control plane that prescribes optimal
pipeline plans, and a data plane that performs
resource reservation-based adaptive batching to handle runtime
dynamics due to asynchronous and bursty request arrivals.
Evaluation results on production workloads show that \name achieves
32.2\%--75.1\% higher serving throughput compared to various baselines.
In future work, we will explore extending \name to support transformer-based
models, for instance, by partitioning them
across high- and low-class GPUs. This would improve
the utilization of low-class GPUs and enhance the overall serving throughput,
while still meeting SLO requirements.

\section*{Acknowledgments}
We thank the anonymous reviewers and our shepherd Purushottam (Puru) Kulkarni
for their helpful comments.
This work is supported in part by NSF grants 2112778, 2211459,
and 2415216.

\newpage
\bibliographystyle{plain}
\bibliography{main}

\appendix
\clearpage

\renewcommand{\thesubsection}{A.\arabic{subsection}}

\section*{Appendix}

\subsection{Mathematical Representation of Basic MILP Formulation}
\label{sec:milp}

As shown in \autoref{tab:milp-input}, the MILP formulation takes as input the
cluster configuration, the latency SLO, and profiling information of the target
DNN model. For the convenience of formulating mathematical constraints, the raw
inputs are further processed and transformed into different representations.
\autoref{tab:milp-var} lists both output decision variables and intermediate
decision variables that will only be used inside the MILP formulation. The MILP
solution outputs the partition points for each pipeline, as well as the batch
size and number of GPUs used by each partition. The MILP formulation maximizes
the total inference throughput across all pipelines in the cluster:
\begin{align}
	\text{maximize} \quad \textstyle\sum_l x_l \label{eq:maximize}
\end{align}

The optimization is under the constraint that inference latency of each
pipeline does not exceed the latency SLO, and the total number of GPUs
allocated across partitions does not exceed the cluster configuration. The
constraints are formally formulated below.

\begin{table}[h]
	\caption{Inputs to the MILP formulation (top) and values derived from inputs (bottom).}
	\label{tab:milp-input}
	\small
	\begin{tabularx}{\columnwidth}{cX}
		\toprule
		Input & \multicolumn{1}{c}{Description} \\
		\midrule
		$N_k$ & GPU count of GPU class~$k$ \\
		$T$ & The latency SLO \\
		$L_{kbi}$ & The inference latency of layer~$i$ under batch size~$b$ on GPU class~$k$ \\
		$S_i$ & The output feature map size of layer~$i$ under batch size~1 \\
		\midrule
		$D_l$ & Number of partitions in pipeline $l$ \\
		$G_k$ & A list of tuples $(l, d)$ indicating GPU class~$k$ is used for partition $d$ of pipeline $l$ \\
		$M$ & Number of layers in the DNN model \\
		$C_{ldbij}$ & The inference latency of DNN partition consisting of layers $i$ to $j$ (exclusive) with batch size $b$ on GPU class associated with $(l, d)$ \\
		$X_{ldbij}$ & The inference throughput of DNN partition consisting of layers $i$ to $j$ (exclusive) with batch size $b$ on a single GPU associated with $(l, d)$ \\
		$Y_{bj}$ & The transfer latency of the feature map of layer $j -1$ with batch size $b$ \\
		\bottomrule
	\end{tabularx}
\end{table}

\begin{table}
	\caption{Output (top) and intermediate (bottom) decision variables in the MILP formulation.}
	\label{tab:milp-var}
	\small
	\begin{tabularx}{\columnwidth}{cX}
		\toprule
		Variable & \multicolumn{1}{c}{Description} \\
		\midrule
		$p_{ldbij} \in \{0, 1\}$ & Whether partition~$d$ in pipeline~$l$ spans from layer $i$ to $j$ (exclusive) and runs at batch size~$b$ \\
		$g_{ldbij} \in \mathbb{N}$ & Number of GPUs used by partition~$d$ in pipeline~$l$ \\
		\midrule
		$t_{ld} \in \mathbb{R}_{\ge 0}$ & Inference latency of partition~$d$ in pipeline~$l$ \\
		$x_{ld} \in \mathbb{R}_{\ge 0}$ & Inference throughput of partition~$d$ in pipeline~$l$ \\
		$n_{ld} \in \mathbb{R}_{\ge 0}$ & Transfer latency between partition~$d$ and partition~$d+1$ in pipeline $l$ \\
		$x_{l} \in \mathbb{R}_{\ge 0}$ & Inference throughput of pipeline~$l$ \\
		\bottomrule
	\end{tabularx}
\end{table}

{\small
\begin{align}
\textstyle \sum_{bij} p_{ldbij} &= 1 && \forall l, d \label{eq:1.1} \\
p_{ldbij} &= 0 && \forall l, d, b, i \ge j \\
\textstyle \sum_{bi} p_{ldbij} = 1 &\rightarrow \textstyle \sum_{b'j'} p_{ld'b'i'j'} = 1 && \forall l, d' = d + 1, i' = j \\
\textstyle \sum_{bj} p_{ldbij} &= 1 && \forall l, d = 0, i = 0 \\
\textstyle \sum_{bi} p_{ldbij} &= 1 && \forall l, d = D_l - 1, j = M \label{eq:1.5} \\
p_{ldbij} = 0 &\rightarrow g_{ldbij} = 0 && \forall l, d, b, i, j \\
p_{ldbij} = 1 &\rightarrow g_{ldbij} \ge 1 && \forall l, d, b, i, j \\
\textstyle \sum_{bij,(l,d)\in G_k} g_{ldbij} &\le N_k && \forall k \label{eq:1.8} \\
t_{ld} &= \textstyle \sum_{bij} C_{ldbij} \cdot p_{ldbij} && \forall l, d \label{eq:1.9} \\
x_{ld} &= \textstyle \sum_{bij} X_{ldbij} \cdot g_{ldbij} && \forall l, d \\
n_{ld} &= \textstyle \sum_{bij} Y_{bj} \cdot p_{ldbij} && \forall l, d \label{eq:1.11} \\
\textstyle \sum_d t_{ld} + \textstyle \sum_d n_{ld} &\le T && \forall l \label{eq:1.12} \\
x_l &= \textstyle \min_d x_{ld} && \forall l
\end{align}
}

Equations~\eqref{eq:1.1}--\eqref{eq:1.5} ensure DNN partitions are well formed,
\ie partitions cannot be empty, the last and first layers in adjacent
partitions must also be adjacent, and the first partition within a pipeline
must start with the first layer, while the opposite applies to the last
partition. Equation~\eqref{eq:1.8} represents the constraint on the total GPU
count. Equations~\eqref{eq:1.9}--\eqref{eq:1.11} calculates for each partition
the inference latency, inference throughput, and transfer latency,
respectively. Finally, Equation~\eqref{eq:1.12} enforces the latency SLO
constraint.

The formulation can be easily scaled to the case of multiple DNN models, where
each DNN model has its own set of decision variables and constraints, with the
additional constraint that the total number of GPUs allocated to all DNN models
does not exceed the cluster configuration.

\clearpage
\subsection{Mathematical Representation of MILP Formulation with Batch Size Unification}
\label{sec:milp-mps}

\begin{table}[H]
    \caption{Inputs to the MILP formulation (with batch size unification) (top) and values derived from inputs (bottom).}
	\label{tab:milp-mps-input}
	\small
	\begin{tabularx}{\columnwidth}{cX}
		\toprule
		Input & \multicolumn{1}{c}{Description} \\
		\midrule
		$N_k$ & GPU count of GPU class~$k$ \\
		$T$ & The latency SLO \\
		$L_{kvbi}$ & The inference latency of block~$i$ under batch size~$b$ on virtual GPU of size $1/v$ and GPU class~$k$ \\
		$S_i$ & The output feature map size of block~$i$ under batch size~1 \\
		\midrule
		$D_l$ & Number of blocks in pipeline $l$ \\
		$G_k$ & A list of tuples $(l, d)$ indicating GPU class~$k$ is used for partition $d$ of pipeline $l$ \\
		$M$ & Number of layers in the DNN model \\
		$C_{ldvbij}$ & The inference latency of DNN partition consisting of blocks $i$ to $j$ (exclusive) with batch size $b$ on $1/v$ virtual GPU of GPU class associated with $(l, d)$ \\
		$X_{ldvbij}$ & The inference throughput of DNN partition consisting of blocks $i$ to $j$ (exclusive) with batch size $b$ on $1/v$ virtual GPU of GPU class associated with $(l, d)$ \\
		$Y_{bj}$ & The transfer latency of the feature map of block $j -1$ with batch size $b$ \\
		\bottomrule
	\end{tabularx}
\end{table}

\begin{table}[H]
    \caption{Output (top) and intermediate (bottom) decision variables in the MILP formulation with batch size unification.}
	\label{tab:milp-mps-var}
	\small
	\begin{tabularx}{\columnwidth}{cX}
		\toprule
		Variable & \multicolumn{1}{c}{Description} \\
		\midrule
		$p_{ldvbij} \in \{0, 1\}$ & Whether partition~$d$ in pipeline~$l$ spans from block $i$ to $j$ (exclusive) and runs at batch size~$b$ on $1/v$ virtual GPU \\
		$g_{ldvbij} \in \mathbb{N}$ & Number of virtual GPUs used by partition~$d$ in pipeline~$l$ \\
		\midrule
		$t_{ld} \in \mathbb{R}{\ge 0}$ & Inference latency of partition~$d$ in pipeline~$l$ \\
		$x_{ld} \in \mathbb{R}_{\ge 0}$ & Inference throughput of partition~$d$ in pipeline~$l$ \\
		$n_{ld} \in \mathbb{R}_{\ge 0}$ & Transfer latency between partition~$d$ and partition~$d+1$ in pipeline $l$ \\
		$x_{l} \in \mathbb{R}_{\ge 0}$ & Inference throughput of pipeline~$l$ \\
		\bottomrule
	\end{tabularx}
\end{table}

\vspace{10pt}
The inputs and decision variables to the MILP formulation with batch size
unification (\autoref{tab:milp-mps-input} and \autoref{tab:milp-mps-var}) are
similar to that of the basic MILP formulation (\autoref{sec:milp}), with the
exception that both the model profiling inputs and decision variables now
include an additional dimension representing virtual GPUs, and that the
profiling inputs are for blocks instead of layers, and the same applies to the
last two dimensions of the output decision variables. The MILP formulation
optimizes for the same throughput objective:
\begin{align}
	\text{maximize} \quad \textstyle\sum_l x_l \label{eq:maximize}
\end{align}

The optimization is under a similar set of constraints as shown below.

{\small
\begin{align}
\textstyle \sum_{vbij} p_{ldvbij} &= 1 && \forall l, d \\
p_{ldvbij} &= 0 && \forall l, d, v, b, i \ge j \\
\textstyle \sum_{vi} p_{ldvbij} = 1 &\rightarrow \textstyle \sum_{v'j'} p_{ld'v'bi'j'} = 1 \notag \\
&\phantom{\rightarrow {}} \forall l, b,\ d' = d + 1,\ i' = j \label{eq:3.3} \\
\textstyle \sum_{vbj} p_{ldvbij} &= 1 && \forall l, d = 0, i = 0 \\
\textstyle \sum_{vbi} p_{ldvbij} &= 1 && \notag \\
&\phantom{\rightarrow {}} \forall l, d = D_l - 1, j = M \\
p_{ldvbij} = 0 &\rightarrow g_{ldvbij} = 0 && \forall l, d, v, b, i, j \\
p_{ldvbij} = 1 &\rightarrow g_{ldvbij} \ge 1 && \forall l, d, v, b, i, j \\
\textstyle \sum_{vbij,(l,d)\in G_k} g_{ldvbij} / v &\le N_k && \forall k \label{eq:3.8} \\
t_{ld} &= \textstyle \sum_{vbij} C_{ldvbij} \cdot p_{ldvbij} && \forall l, d \\
x_{ld} &= \textstyle \sum_{vbij} X_{ldvbij} \cdot g_{ldvbij} && \forall l, d \\
n_{ld} &= \textstyle \sum_{vbij} Y_{bj} \cdot p_{ldvbij} && \forall l, d \\
\textstyle \sum_d t_{ld} + \textstyle \sum_d n_{ld} &\le T && \forall l \\
x_l &= \textstyle \min_d x_{ld} && \forall l
\end{align}
}

The major differences lie in Equations~\eqref{eq:3.3} and \eqref{eq:3.8}. In
Equations~\eqref{eq:3.3}, the dimension $b$ is not summed over, which enforces
that the same batch size used by the first partition will also need to be used
by the next partition. In Equations~\eqref{eq:3.8}, the decision variable $g$
represents the count of \textit{virtual} GPUs, thus, dividing it by $v$ gives
us the number of \textit{physical} GPUs that the partition uses (note that $v$
is not a decision variable and such divisions are allowed in MILP).

\clearpage

\subsection{The Resource Reservation-Based Adaptive Batching Algorithm}
\label{subsec:batch}

\autoref{alg:batch} shows the pseudocode for the resource reservation-based
adaptive batching algorithm, which was explained in \autoref{subsec:sched}.
To recap, the algorithm first identifies a pooled pipeline that can complete a
batched inference at the pipeline's unified batch size $bs_i$ with the shortest
waiting time (lines 4--8).
It then searches for the largest batch size that can satisfy the SLO (lines
9--14).

The two supporting functions used by \autoref{subsec:sched}, \texttt{probe()}
and \texttt{reserve()}, is shown in \autoref{alg:reserve}.
The helper function \texttt{earliestSlot(res, t, l)} returns the earliest time
(no earlier than $t$) when a list of resources \texttt{res} are free for
duration $l$.
First, since feature map transfer
requires the network resources on both sending and receiving sides to
be available at the same time,
we use \texttt{earliestSlot(res, t, l)} to find the earlist available transfer
slot that works for both the last GPU's uplink and current GPU's downlink
(lines 12--14). After reserving the two links for feature map transfer,
we update current time $t$ and then find and reserve
the earliest available inference time slot for the current GPU (lines 15--16).

\begin{algorithm}[h]
	\small
	\SetKw{KwTo}{in}
	\SetKw{KwDownTo}{down to}
	\SetKw{KwBreak}{break}

    \textbf{Inputs}: $q$ (pending requests sorted by arrival)\;
    \hspace{1cm} $pooled\_pipelines$ (pooled pipelines in the cluster)\;

	\While{true}{
		\tcp{choose the pooled pipeline}
		$p^*=nil, t^* = \infty$\;
		\For{$p$ \KwTo $pooled\_pipelines$}{
			$r$ = probe($p$, $p$.bs)\;
			\If{waitTime($r$) < $t^*$}{
                $t^*$ = waitTime($r$), $p^*$ = $p$\;
			}
		}
		\BlankLine
		\tcp{choose the pipeline path and batch size}
		$bs^*=nil, path^*=nil, resv^*=nil$\;
		\For{$bs$ = $p^*$.bs \KwDownTo 0}{
			$path, resv$ = probe($p^*$, $bs$)\;
			\If{finishTime($resv$) $\le q$[0].deadline}{
		        $bs^*=bs, path^*=path, resv^*=resv$\;
				\KwBreak\;
			}
		}
		\BlankLine
		\tcp{perform request drop, wait, or dispatch}
		\uIf{bs == 0}{
			drop $q$[0]\;
		}
		\uElseIf{$q$.length < bs}{
			wait for more requests till requests in $q$ are about to miss deadline\;
		}
		\Else{
			reserve($r$)\;
			dispatch first bs requests in $q$ according to $r$\;
		}
	}

	\caption{Resource reservation-based adaptive batching.}
	\label{alg:batch}
\end{algorithm}

\newpage

\begin{algorithm}
	\small
	\SetKwProg{Fn}{function}{}{}
	\SetKw{KwTo}{in}
	\SetKw{KwNot}{not}

	\Fn{probe($pooled\_pipeline$, $bs$)}{
        \textbf{Inputs}: The assumed selected pooled pipeline and the batch size\;
        \textbf{Outputs}: The optimal pipeline path and the resources that need to be reserved\;
        $t_g$ = now(), $path$ = [], $resv$ = []\;
		\For{partition \KwTo $pooled\_pipeline$}{
			$l_n$ = calcNetLat(partition, $bs$)\;
			$l_i$ = calcInferenceLat(partition, $bs$)\;
			\BlankLine
			$t^* = \infty$, $r^*$ = []\;
			\For{gpu \KwTo partition}{
				$t = t_g$, $r$ = []\;
				\tcp{est.\ time to transfer feature map}
				\If{\KwNot first partition}{
					$u$ = lastGpu.netUL, $d$ = gpu.netDL\;
					$t$ = earliestSlot([$u$, $d$], $t$, $l_n$)\;
					$r$ += [\{$u$, $t$, $l_n$\}, \{$d$, $t$, $l_n$\}], $t$ += $l_n$\;
				}
				\BlankLine
				\tcp{est.\ time to finish inference}
				$t$ = earliestSlot([gpu], $t$, $l_i$)\;
				$r$ += [\{gpu, $t$, $l_i$\}], $t$ += $l_i$\;
				\If{$t < t^*$}{
					$t^* = t$, $r^* = r$, gpu$^*$ = gpu\;
				}
			}
			$t_g = t^*$, lastGpu = gpu$^*$\;
            $path$ += gpu$^*$, $resv$ += $r^*$\;
		}
		\KwRet{($path$, $resv$)}\tcp*[l]{resource usage}
	}
	\BlankLine
	\Fn{reserve($resv$)}{
		\For{\{res, start, dur\} \KwTo $resv$}{
			markReserved(res, start, dur)\;
		}
	}

	\caption{Resource reservation functions.}
	\label{alg:reserve}
\end{algorithm}

\end{document}